\documentclass[iop]{emulateapj}

\usepackage{textcomp}
\usepackage{array,multirow}
\usepackage{natbib}
\defcitealias{KH09}{KH09}

\begin{document}

\title{The Intrinsic Eddington Ratio Distribution of Active Galactic Nuclei in Star-Forming Galaxies from the Sloan Digital Sky Survey}

\author{Mackenzie L. Jones\altaffilmark{1}, Ryan C. Hickox\altaffilmark{1}, Christine S. Black\altaffilmark{1}, Kevin N. Hainline\altaffilmark{1,2}, Michael A. DiPompeo\altaffilmark{1}, Andy D. Goulding\altaffilmark{3}}
\affil{$^{1}$Department of Physics and Astronomy, Dartmouth College, Hanover, NH 03755}
\affil{$^{2}$Steward Observatory, University of Arizona, Tucson, AZ 85721}
\affil{$^{3}$Department of Astrophysical Sciences, Princeton University, Princeton, NJ 08544, USA}

%%%%%%%%%%%%%%%%%%%%%%%%%%%%%%%%%%%%%%%%%%%%%%%%%
% ABSTRACT
%%%%%%%%%%%%%%%%%%%%%%%%%%%%%%%%%%%%%%%%%%%%%%%%%
\begin{abstract}
An important question in extragalactic astronomy concerns the distribution of black hole accretion rates of active galactic nuclei (AGN). Based on observations at X-ray wavelengths, the observed Eddington ratio distribution appears as a power law, while optical studies have often yielded a lognormal distribution. There is increasing evidence that these observed discrepancies may be due to contamination by star formation and other selection effects. Using a sample of galaxies from the Sloan Digital Sky Survey Data Release 7, we test if an intrinsic Eddington ratio distribution that takes the form of a Schechter function is consistent with previous work that suggests that young galaxies in optical surveys have an observed lognormal Eddington ratio distribution. We simulate the optical emission line properties of a population of galaxies and AGN using a broad instantaneous luminosity distribution described by a Schechter function near the Eddington limit. This simulated AGN population is then compared to observed galaxies via the positions on an emission line excitation diagram and Eddington ratio distributions. We present an improved method for extracting the AGN distribution using BPT diagnostics that allows us to probe over one order of magnitude lower in Eddington ratio counteracting the effects of dilution by star formation. We conclude that for optically selected AGN in young galaxies, the intrinsic Eddington ratio distribution is consistent with a possibly universal, broad power law with an exponential cutoff, as this distribution is observed in old optically selected galaxies and in X-rays.
\end{abstract}


%%%%%%%%%%%%%%%%%%%%%%%%%%%%%%%%%%%%%%%%%%%%%%%%%
% INTRODUCTION
%%%%%%%%%%%%%%%%%%%%%%%%%%%%%%%%%%%%%%%%%%%%%%%%%
\section{Introduction}\label{sec:int}

In the past decade, astronomers have made significant progress in developing a generalized model of the formation and evolution of galaxies, their host bulges, and black holes across cosmic time (For reviews see \citealt{Sil12,Ale12}). Despite this progress, the evolution of supermassive black holes (SMBH) and their impact on their host galaxies and large-scale structures remains relatively poorly understood. As these central black holes grow via mass accretion they emit copious amounts of light as active galactic nuclei (AGNs; \citealt{Sal64,Lyn69,Sha73,Sol82,Ree84}). The masses of these SMBHs are correlated with the properties of their host stellar bulges \citep{Kor13} and the volume-average galaxy-black hole growth rate is consistent with the black hole-spheroid mass relationship for a wide range of black hole masses (e.g., \citealt{Hec04}). However, other than a potential link through a common supply of cold gas, the physical processes connecting black holes and galaxies are still uncertain \citep{Ale12}.

AGN are observed and characterized differently in various wavelength regimes (e.g., \citealt{Hic09,Men13,Gou14}).  Decades of optical observations of AGN have led to the characterization of AGN based on spectroscopic characteristics (e.g., \citealt{Pet97}). Particularly useful for high-z studies due to their brightness in the optical, type 1 AGN exhibit optical spectra containing blue continua and both narrow and broad line emission, while the spectra of type 2 AGN contain only narrow line emission (For reviews see \citealt{Ant93,Urr95,Net15}). The most statistically powerful studies of AGN to date come from optical spectroscopic surveys such as the Sloan Digital Sky Survey (SDSS; \citealt{Yor00}). In surveys of type 2 AGN, it is possible to connect accretion to the galaxy properties via high excitation narrow lines without contamination from the optical continuum that is characteristic of type 1 AGN. Narrow line AGN may be selected based on specific emission line ratio strengths (e.g.,\citealt{Vei87,Kew01,Kau03,Kew06}) and the corresponding positions on the \citet{BPT81} BPT diagram.

Likewise, in the infrared, specific emission line ratio strengths can be used to select AGN (e.g., \citealt{Pop08,Gou09,Pet11}). More commonly, IR-selected AGN are characterized photometrically using color selection (e.g., \citealt{Lac07,Jar11,Don12,Ste12,Mat12}). In the X-rays, AGN appear as the most luminous X-ray sources and can be selected based on an X-ray luminosity threshold and a characteristic X-ray power-law spectrum (e.g., \textit{Chandra Deep Fields}; \citealt{Bra01,Gia02,Ale03,Luo08,Luo10}).

Once AGN are selected through observations, measuring the distribution of the Eddington ratio (a ratio of the observed accretion rate to the maximum possible accretion given by the Eddington limit) yields important information about black hole growth, such as the typical accretion rate for black holes of a given mass, and provides a clue into the physical properties driving this growth (e.g., \citealt{Hec04,Air12}). Individual galaxies can vary dramatically in AGN luminosity over short timescales (e.g., \citealt{Sch10,Hic14,Sch15var}) 
so measuring the statistical distribution of Eddington ratios can provide a general framework for the long-term process of black hole growth. 

%%%%%%%%%%%%%%%%%%%%%%%%%%%%%%%%%%%%%%%%%%%%%%%%%
\begin{figure}[!t]
\resizebox{80mm}{!}{\includegraphics{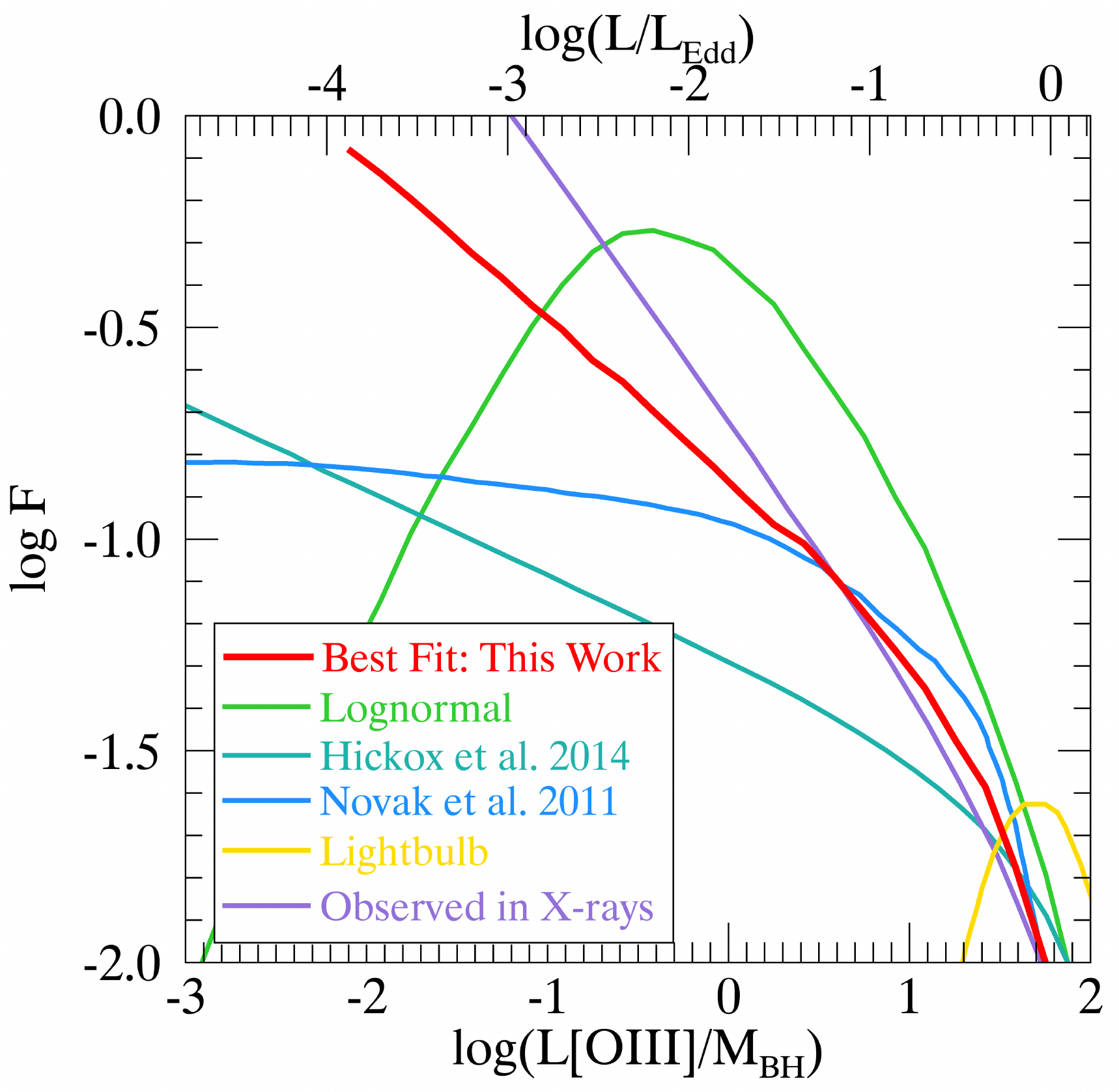}} \\
\caption{A variety of intrinsic Eddington ratio distributions are used in theoretical models and simulations such as varieties of Schechter and lognormal functions, including the roughly lognormal distribution derived using the method of \citetalias{KH09} (green; refer to Section \ref{sec:khreal}). Our chosen best fit model of the intrinsic Eddington ratio distribution (red; refer to Section \ref{sec:khsim}) is shown compared to previously published distributions.\label{fig:alledd}}
\end{figure}
%%%%%%%%%%%%%%%%%%%%%%%%%%%%%%%%%%%%%%%%%%%%%%%%%

Many theoretical models and simulations predict an intrinsic Eddington ratio distribution (\citealt{Nov11, Gab13,Tha14}) or assume a functional form for the Eddington ratio distribution in modeling the AGN population (Figure \ref{fig:alledd}). More complicated hydrodynamical models can directly yield the intrinsic Eddington ratio distribution, while many analytic or semi-analytic models must assume a distribution because they don't have the physical capability to predict one. The distributions used in these models vary; some theoretical studies utilize a lognormal shape. One particular subset of the lognormal distribution is often referred to as the ``scattered lightbulb,'' in which the AGN accretion is either ``on'' or ``off'' \citep{Con13}. Other simulations choose a Schechter function, a power law with an exponential cutoff at high Eddington ratios which is often generalized at low Eddington as a power law, (e.g., \citealt{Hop09Her}, in which they argue against the ``lightbulb'' model). \citet{Hop09} also uses a Schechter function for the intrinsic AGN distribution:

%%%%%%%%%%%%%%%%%%%%%%%%%%%%%%%%%%%%%%%%%%%%%%%%%
\begin{equation}\label{equ:fidu}
\frac{dt}{d\log L}=t_0 \left(\frac{L}{L_{cut}}\right)^{-\alpha}\exp\left(-L/L_{\rm{cut}}\right).
\end{equation}
%%%%%%%%%%%%%%%%%%%%%%%%%%%%%%%%%%%%%%%%%%%%%%%%%

In addition to modeling, the quasar luminosity function (QLF) is another probe into the intrinsic Eddington ratio distribution. Building on the work of \citet{Con13} in which the QLF is fit with a ``lightbulb'' model, \citet{Vea14} compares the lognormal shape to a truncated power law in a simulated fit to the QLF. They find that at high luminosities both distributions fit the QLF, but at low luminosities the fits are poorly constrained, which they interpret as evidence that quasars spend a majority of their time at low luminosities.

To further study the intrinsic AGN Eddington ratio distribution, it is useful to examine the connection between the SMBH and host galaxy. There is increasing observational evidence that the average black hole accretion rate is related to host galaxy star formation rate in AGN (e.g., \citealt{Che13}), whereas for individual galaxies this dependency is observed to be much weaker (e.g., \citealt{Sha10,Ros12,Ale12}). This relationship can be reproduced with a tight correlation between star formation and time-averaged AGN activity, along with short-term variability of the AGN over a wide dynamic range (\citealt{Tha14,Vol15}). \citet{Hic14} finds that a Schechter function AGN luminosity distribution can reproduce observed trends.

Direct measurements can also probe the Eddington ratio distribution of galaxies. In the X-rays, which are less susceptible than the optical to extinction and contamination biases (as discussed below), \citet{Air12} and \citet{Bon12} find an observed Eddington ratio distribution that is similar to the Schechter function. The Eddington ratio distribution of quasars can also be observed directly using optical spectroscopy; e.g., \citet{Kol06} and \citet{She08} find for luminous AGN, the Eddington ratio distribution is lognormal and peaks at high Eddington, while low luminosity AGN can have distributions that extend to lower Eddington ratios \citep{Gav08}. \citet{Scz10} observe a turnover at low Eddington ratios for low redshift AGN in the Hamburg/ESO survey but attribute this to selection effects and determine that an intrinsic Schechter function better describes the Eddington ratio distribution. There is further evidence that the Eddington ratio distribution evolves with redshift which may be indicative of a shift in the global black hole accretion density to lower black hole masses at lower redshift (e.g. \citealt{Nob12,She12}).

\citet[][hereafter KH09]{KH09} present a method for determining the AGN Eddington distribution from optical spectroscopic galaxy samples, selecting AGN based on the position of galaxies with narrow emission lines on the BPT diagram. They determine the distribution of AGN Eddington ratios for different galaxy populations and test how these distributions vary as a function of black hole mass. An outline of this method is provided in Section \ref{sec:khreal}. Two distinct AGN populations in different host galaxies are separated based on a mean stellar age indicator, the relative strength change of continuum flux across the 4000 Angstrom break \citep{Bal99}. A young stellar population is defined as having a break index of D4000$<$1.5 and an old stellar population has a break index of D4000$>$1.5 \citep{Kau03}. The Eddington ratio distribution for the young galaxy population is found to be lognormal in shape, while the old galaxy population has a shape consistent with a power law.

Both the lognormal and power law models have successfully been used to fit the Eddington ratio distribution in various studies. In order to fully compare these Eddington ratio distributions, it is necessary to understand any observational biases that could be intrinsic to the methods. 

While X-ray observations provide one of the most valuable methods for detecting and characterizing AGN, studies of the cosmic X-ray background (CXB) and direct X-ray measurements (e.g., \citealt{Ste14,Lan14}) suggest that X-ray observations are biased against the most heavily obscured AGN. The CXB is known to be produced primarily by accretion, the majority of which is AGN accretion while a smaller contribution is due to X-ray binaries. This background radiation peaks in the hard X-rays ($\sim$30 keV). However, observations have shown that the majority of AGN are identified in the soft X-rays ($0.1-2$ keV; e.g., \citealt{Mar80}) and do not produce enough hard X-rays to account for the full hard CXB, implying that we are missing a significant fraction of AGN due to obscuration (e.g., \citealt{Hop06,Cle02,Spo04,Iwa05,Dow07}).

Like the X-ray, optical methods suffer from selection effects. Dilution from host galaxy light and obscuration of the AGN can affect optical emission lines, either contributing to their flux or causing them to disappear altogether (\citealt{Hop09,Gou09}). Similarly, AGN emission in the IR is also subject to dilution from star formation (\citealt{Pol07,Mul11,Mul12}). Recent work by \citet{Tru15} finds a strong bias for optical line-ratio selected AGN at a given accretion rate, such that for a fixed Eddington ratio, fewer AGN are identified in low mass, star-forming host galaxies than in massive and low star-forming hosts. It has also been shown that dilution in spectroscopic samples depends significantly on the size of the aperture used to extract the flux \citep{Mar14}. Additionally, aperture effects can cause extended emission from stellar processes to mimic the emission of a LINER (e.g., \citealt{Yan12}).

In light of these selection effects, this paper aims to determine the intrinsic Eddington ratio distribution for type 2 AGN in SDSS galaxies. Specifically, we will test whether an intrinsic Schechter function  (e.g., \citealt{Hop09,Che13,Hic14}) is consistent with the observational results obtained by \citetalias{KH09} for a young galaxy population. We simulate a galaxy population with a known intrinsic AGN Eddington ratio distribution, where the distribution acts as a proxy for the amount of time an AGN spends at a particular Eddington ratio, and apply the method of \citetalias{KH09} to determine the ``observed'' AGN Eddington ratio distribution. We present the observations used in Section \ref{sec:obs} and the application of the \citetalias{KH09} method to observed data in Section \ref{sec:khreal}. Our fiducial model is defined in Section \ref{sec:mod}, and the description and results of the application of the \citetalias{KH09} method to our simulated data is presented in Section \ref{sec:khsim}. Furthermore, in Section \ref{sec:frac} we present a new method for extracting the AGN contribution to the flux using optical emission line ratios. A discussion and summary are given in Section \ref{sec:dis}.

%%%%%%%%%%%%%%%%%%%%%%%%%%%%%%%%%%%%%%%%%%%%%%%%%
% DATA
%%%%%%%%%%%%%%%%%%%%%%%%%%%%%%%%%%%%%%%%%%%%%%%%%
\section{Data}\label{sec:obs}

Our galaxy sample originates from the SDSS Data Release 7 \citep{Yor00,Str02,Aba09}. We obtained spectroscopic information from the MPA-JHU value-added catalog\footnote{http://www.mpa-garching.mpg.de/SDSS/DR7/}. All emission line fluxes were previously corrected for Galactic reddening within the catalog following \citet{Odo94}. The fluxes were extinction corrected via the \citet{Cha00} prescription, following \citetalias{KH09}. The extinction corrected [\ion{O}{3}] fluxes then underwent a mean bolometric correction of 600, as selected by \citetalias{KH09}. Star formation rates were available in the catalog and were determined following the method of \citet{Bri04}. We calculated black hole masses via stellar velocity dispersions and the formula given by \citet{Tre02}. The available stellar mass estimates in the value added catalog were determined based on photometric fits with stellar population synthesis models (\citealt{Kau03,Sal07}). The chosen sample has a spectroscopic redshift range of 0.0$<$z$<$0.33. After the extinction and bolometric flux corrections, we select our ``real'' sample to be all observed SDSS objects with a stellar mass measurement. 

%%%%%%%%%%%%%%%%%%%%%%%%%%%%%%%%%%%%%%%%%%%%%%%%%
% REAL EDDINGTON
%%%%%%%%%%%%%%%%%%%%%%%%%%%%%%%%%%%%%%%%%%%%%%%%%
\section{Determining the Eddington Ratio Distribution: Observed SDSS Galaxies}\label{sec:khreal}

To determine the observed Eddington ratio distribution, we first separate our ``real'' sample of SDSS observed galaxies into two populations by mean stellar age, as defined by the D4000 index. The observed young and old galaxy samples are defined as having a D4000 break below and above 1.5, respectively. A break index of 1.5 is chosen based on the results of \citetalias{KH09}, in which they find that the observed Eddington ratio distribution showed little dependence on D4000 for all galaxies with D4000$<$1.5. Furthermore, we use this simple definition for our ``young'' sample to include as large as possible a sample in our analysis, rather than the \citetalias{KH09} representative ranges that separate the populations into young and old by a D4000 break below 1.4 and above 1.7, respectively. We can then further split the galaxies into groups based on the strength of the emission signal to noise (S/N).  We define ``[\ion{O}{3}] undetected'' galaxies as having a S/N$<$3 in [\ion{O}{3}] while galaxies that are ``[\ion{O}{3}] detected'' are defined as having S/N$>3$ in [\ion{O}{3}]. Of the $\sim$770,000 SDSS galaxies retrieved from the MPA-JHU value added catalog with mass estimates, there are $\sim$520,000 that are classified as [\ion{O}{3}] detected while $\sim$250,000 are [\ion{O}{3}] undetected. Of the 
[\ion{O}{3}] detected galaxies, $\sim$306,000 of these fit our young classification.

We focus on the young sample in our analysis, in part because these galaxies are actively star-forming and suffer greatly from contamination and other selection effects. In addition, previous optical studies have yielded a lognormal Eddington ratio distribution for young galaxies, while the old galaxy population exhibits an Eddington ratio distribution that is consistent with the power law distribution found with X-ray observations.

With the young galaxy population taken from SDSS DR7, we next reproduce the \citetalias{KH09} prescription for determining the Eddington ratio distribution. Of our real sample of [\ion{O}{3}] detected young galaxies, we further select galaxies with significant line detection (S/N$>3$) in all four emission lines used for the BPT diagram (H$\alpha$, H$\beta$, [\ion{N}{2}], and [\ion{O}{3}]) to directly compare our results with those in \citetalias{KH09}. Selecting galaxies with a detection in all four emission lines rather than our original cut in [\ion{O}{3}] results in a decrease of roughly $4\%$ of our initial sample. We thus select our ``real'' observed sample to consist of 294,934 young galaxies with significant line detections.

\begin{figure}[!t]
\resizebox{80mm}{!}{\includegraphics{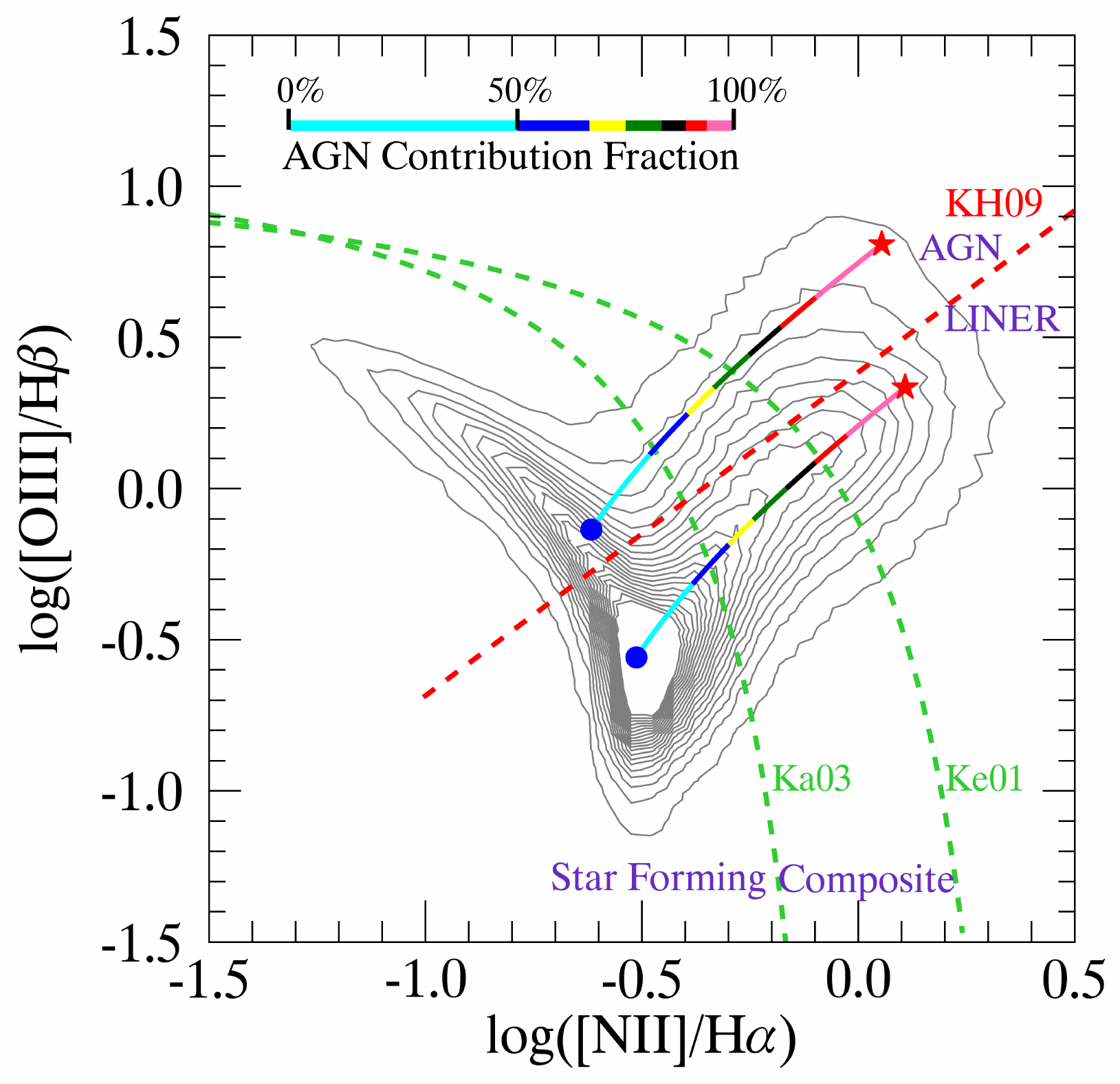}} \\
\caption{BPT diagram of the full sample of observed SDSS DR7 galaxies with the \citetalias{KH09} tracks used to determine the AGN contribution based on distance from the ``pure'' SF sequence (blue circle) to the ``pure'' AGN region (red star). The AGN contribution fractions follow the color coding of \citetalias{KH09}. \label{fig:rtrack}}
\end{figure}

Following the \citetalias{KH09} analysis, we determine the AGN contribution to the total [\ion{O}{3}] flux using the BPT diagram (Figure \ref{fig:rtrack}). An assumption is made that galaxies defined at a particular locus on the star-forming sequence have no AGN contribution, while galaxies at a locus on the far end of the AGN sequence have line emission entirely dominated by an AGN. 

%%%%%%%%%%%%%%%%%%%%%%%%%%%%%%%%%%%%%%%%%%%%%%%%%
%%%%%%%%%%%%%%%%%%%%%%%%%%%%%%%%%%%%%%%%%%%%%%%%%

We create tracks of AGN contribution by taking a typical star-forming spectrum (``no AGN'') and adding increasing emission characteristic of an AGN spectrum until the emission line ratios on the BPT diagram are entirely dominated by an AGN (``pure AGN''). The positions of the tracks on the BPT diagram are used to define a relationship between distance, defined in the space of the log of the two line ratios, from the ``pure AGN'' point and the fractional contribution of the AGN to the total [\ion{O}{3}] emission. Thus for all galaxies we are able to break the total observed [\ion{O}{3}] flux into a star-forming or AGN component based on their individual distances from the ``pure AGN'' track endpoint.

As in \citetalias{KH09}, two tracks have been selected to represent the trajectory for two AGN populations based on their ionization; higher ionization (Seyferts) and lower ionization (LINERs). The chosen sample of SDSS galaxies are split into these two ionization populations via a defined separation line that matches the one used by \citetalias{KH09}. Figure \ref{fig:rtrack} shows these tracks and ionization demarcation line.

The AGN [\ion{O}{3}] contribution to the total flux is used to calculate the Eddington parameter: the AGN [\ion{O}{3}] luminosity divided by the black hole mass, as adopted by \citetalias{KH09}. The fraction of black holes of a given mass per unit logarithmic interval of L[\ion{O}{3}]/M$_{BH}$, or the differential Eddington parameter, represents the distribution of Eddington ratios L/L$_{Edd}$ for AGN.

%%%%%%%%%%%%%%%%%%%%%%%%%%%%%%%%%%%%%%%%%%%%%%%%%
% MODEL
%%%%%%%%%%%%%%%%%%%%%%%%%%%%%%%%%%%%%%%%%%%%%%%%%
\section{Simulating the Galaxy and AGN Population}\label{sec:mod}

Once the Eddington ratio distribution is found for our real sample of observed galaxies, we can proceed in simulating a population with a known intrinsic Eddington ratio distribution. By comparing the output from the \citetalias{KH09} method of our simulated galaxy population with that obtained from our real sample, we may better understand the underlying Eddington ratio distribution for SDSS galaxies. 

\begin{figure}[!t]
\resizebox{80mm}{!}{\includegraphics{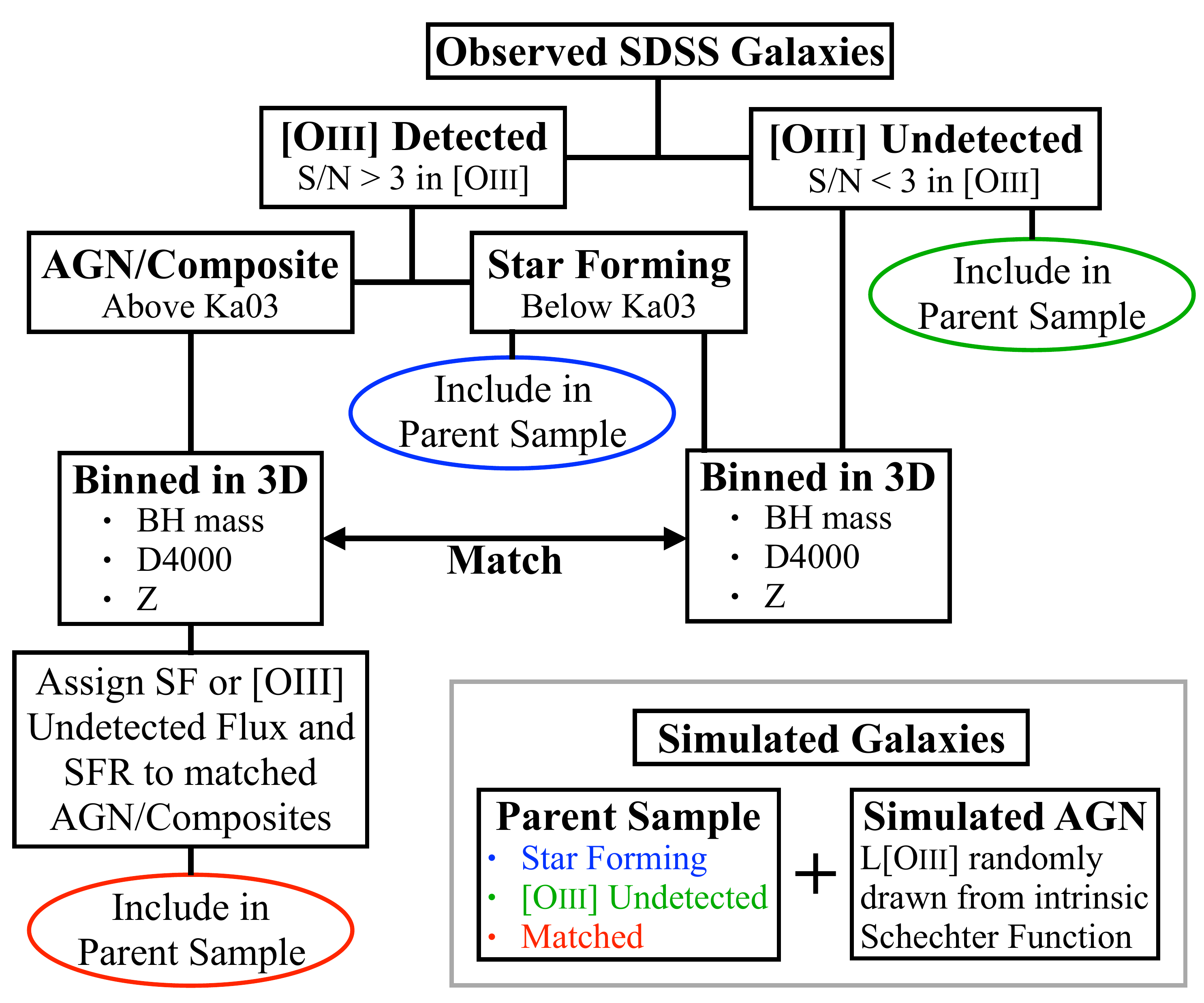}} \\
\caption{Our simulated sample is built from a parent sample of SDSS galaxies and an addition of an AGN component with Eddington ratio drawn randomly from our Schechter function distribution. The parent sample consists of a comparable number of galaxies to our observed SDSS sample by including galaxies with flux that is classified as star-forming on the BPT diagram or [\ion{O}{3}] undetected, or by assigning these fluxes to BPT selected composite and AGN that match in black hole mass, D4000, and redshift. \label{fig:simdia}}
\end{figure}

The simulation is built on the assumption that every galaxy has a supermassive black hole with some accretion based on our intrinsic Eddington ratio distribution and thus, by definition, is an ``AGN''. However, not all of these galaxies will have an accretion rate that is significant enough to be identified using BPT diagnostics and may not be considered an ``observed'' AGN. The simulated sample begins with the SDSS galaxies. A schematic of the selection process in making our simulated sample is outlined in Figure \ref{fig:simdia}. We again separate the SDSS galaxies based on [\ion{O}{3}] detection and age. Objects in the observed sample that have S/N $<$3 in [\ion{O}{3}], corresponding to a median L[\ion{O}{3}] of $7.0\times10^{43}$ erg s$^{-1}$ (z=0.13), are considered to be a part of our [\ion{O}{3}] undetected galaxy class, as similarly defined for our real sample. Those with S/N $>3$ in [\ion{O}{3}] are [\ion{O}{3}] detected and further separated by their position on the BPT diagram. Using the star-forming sequence boundary \citep{Kau03}, the star-forming galaxies are separated from the composite and AGN regime. Using the BPT to separate our [\ion{O}{3}] detected is not strictly accurate for those objects without detections in all four lines, however this accounts for only $4\%$ of our total [\ion{O}{3}] detected sample and furthermore allows for the possibility of those emission lines to become significant with the addition of a simulated AGN component. Our [\ion{O}{3}] undetected galaxy group and star-forming group are then used as a base for our simulated sample.

%%%%%%%%%%%%%%%%%%%%%%%%%%%%%%%%%%%%%%%%%%%%%%%%%
%%%%%%%%%%%%%%%%%%%%%%%%%%%%%%%%%%%%%%%%%%%%%%%%%

In order to better compare with the real observed sample population numbers, we also use objects in the Composite and AGN regime that are [\ion{O}{3}] detected and located on the BPT diagram above the \citet{Kau03} boundary. However, these AGN and Composite galaxies can not be directly included in the parent sample because we have no way of knowing a priori how much the AGN contributes to the total emission line flux values. Instead, they are matched to our earlier pool of [\ion{O}{3}] undetected and star-forming galaxies. This match is run in three dimensions: black hole mass, D4000 line break, and redshift for $\sim$220,000 AGN and Composite galaxies. The distributions of stellar mass and star formation rate for our matched AGN and Composite galaxies are not significantly different than the distributions of the parent sample with which they are matched.

The star-forming and [\ion{O}{3}] undetected galaxies are sorted into a three dimensional cube of black hole mass (bin size of 0.1), D4000 (bin size of 0.02), and redshift (bin size 0.01). The composite and AGN objects are then sorted in the same way. The bins in each of these two 3D cubes are then matched; every composite and AGN galaxy is paired with a randomly selected star-forming or [\ion{O}{3}] undetected galaxy from its matching bin. Those that do not match are then excluded, but this does not amount to a significant decrease in objects ($\sim$7,000, roughly $3\%$). Based on the match, these AGN and Composite galaxies are assigned emission line fluxes and star formation rates of the matched star-forming or [\ion{O}{3}] undetected galaxy before they are included in the parent sample. The black hole mass remains the same in order to match the distribution of the real observed sample, as defined in Section \ref{sec:khreal}. We match to both the [\ion{O}{3}] detected, star-forming galaxies and the [\ion{O}{3}] undetected galaxies, rather than just the [\ion{O}{3}] detected, star-forming galaxies to allow for the possibility that the [\ion{O}{3}] undetected galaxies may become significant with the addition of our simulated AGN component.

Once the simulated AGN component is added to our parent sample (as discussed below) we have a comparable number of simulated galaxies to what is observed. The simulated AGN component is drawn randomly from an Eddington ratio distribution characterized by a Schechter function with an exponential cutoff near the Eddington limit (e.g., \citealt{Hop09,Hic14}), as given by Equation \ref{equ:fidu}. This is then added to the parent sample. We adopt an upper cutoff at the Eddington luminosity ($L_{cut}=L_{edd}=1.38\times10^{38}M_{BH}$ erg s$^{-1}$, assuming a radiative efficiency of $0.1$), while the lower cutoff, which determines the normalization of the curve (i.e. the average luminosity), can be varied to better match the observed distribution on the BPT diagram. To further motivate our specific choice of a Schechter function, we also tested a simple power-law distribution and found the exponential cutoff is needed to match the observed suppression of high Eddington AGN.

\begin{figure}[!t]
\resizebox{80mm}{!}{\includegraphics{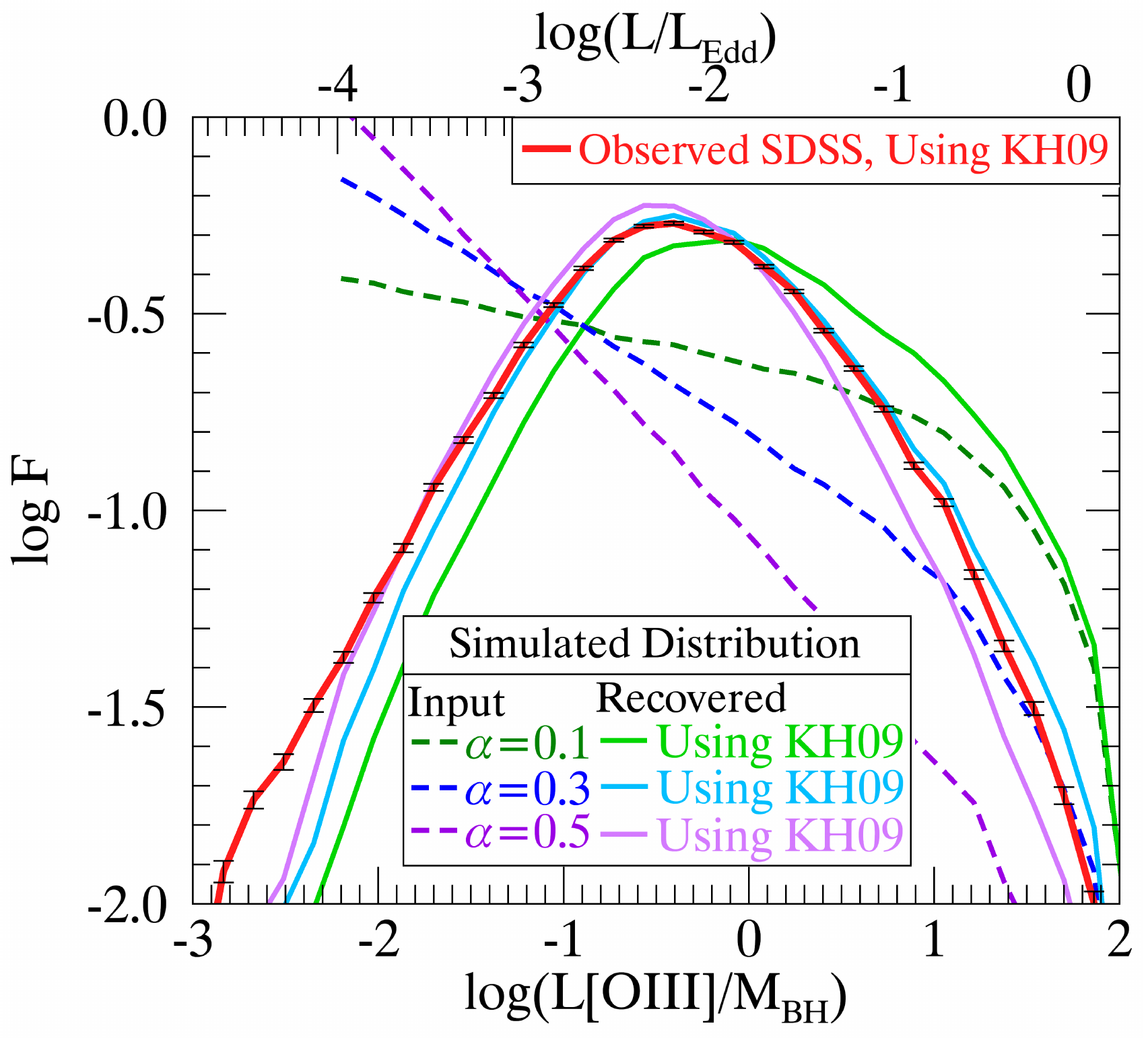}} \\
\caption{An example of the observed Eddington ratio distributions of simulated SDSS galaxies resulting from intrinsic Schechter function Eddington ratio distributions with varying $\alpha$ and a lower cutoff of -4.0 compared to the real observed SDSS Eddington ratio distribution. As demonstrated, changing the slope of the intrinsic function changes the peak ``observed'' distribution and the amplitude of the cutoff. A slope of $\alpha=$ 0.3 most closely resembles the real observed Eddington ratio distribution. The Eddington ratio distributions are shown as the logarithm of the fraction per logarithmic interval of L[\ion{O}{3}]/M$_{BH}$ in solar luminosity and mass. \label{fig:vara}}
\end{figure}

We first vary the slope of the Schechter function, $\alpha$, between 0.0 and 0.8 in steps of 0.1 to determine the best fit to the observed Composite and AGN occupation fractions. From this broad search, we further refine our best fit by varying $\alpha$ between 0.28 and 0.48 in steps of 0.002. Changing the slope of the power law makes the largest impact in the distribution of the most luminous objects, as demonstrated in Figure \ref{fig:vara} by the variation of the Eddington ratio distribution.  In addition to varying $\alpha$ to improve our fit, we also allow the lower cutoff ($\log(L_{cut})$) to vary between -4.8 and -2.8 in steps of 0.02 which changes the number of observed high Eddington ratio galaxies. We use the ratio of L[\ion{O}{3}]/$M_{BH}$ as a proxy for the Eddington ratio using a bolometric correction of 600 \citepalias{KH09} on the instantaneous bolometric luminosities. This corresponds to an addition of 1.78 to the logarithm of the Eddington ratio distribution. For example, our lower cutoffs would then span a range of -3.02 to -1.02 in $\log$(L[\ion{O}{3}]/$M_{BH}$).

%%%%%%%%%%%%%%%%%%%%%%%%%%%%%%%%%%%%%%%%%%%%%%%%%
%%%%%%%%%%%%%%%%%%%%%%%%%%%%%%%%%%%%%%%%%%%%%%%%%

Once we have our simulated L[\ion{O}{3}], we can find the AGN contribution to the other BPT emission lines. Each parent galaxy is assigned an AGN [\ion{O}{3}] luminosity randomly selected from the given Eddington ratio distribution. We then select two regions in BPT space where we could expect an addition of AGN flux to move our parent sample; a high or low ionization circular region with a distribution given as a function of radius (radius = 0.075) and centered near the \citetalias{KH09} AGN track points. Based on the real observed distribution of galaxies in these two regions, we construct a probability for an object in our parent sample to move to either region once an AGN has been added, such that an object moves toward a random point within the high ionization circular region roughly $48\%$ of the time.

We now have an AGN contribution in [\ion{O}{3}], the total emission line ratios [\ion{O}{3}]/H$\alpha$ and [\ion{N}{2}]/H$\beta$ from the randomly selected endpoint, and the observed flux in our four emission lines from the parent sample. By adding the AGN contribution in [\ion{O}{3}] to the observed [\ion{O}{3}] flux and using the total emission ratios, we can determine the total (AGN+SF) emission in each four lines. Since we know the SF component, we can extract the simulated AGN contribution to the remaining emission lines.

By adding these simulated AGN components to the emission from the parent sample in [\ion{O}{3}], [\ion{N}{2}], H$\alpha$, and H$\beta$, we obtain a simulated sample of young galaxies that contain line emission from stellar processes, as well as an AGN component with L$_{bol}$ drawn from our given Schechter function distribution. Thus we have a sample with a known intrinsic Eddington ratio distribution that we can then use to model the unknown intrinsic Eddington ratio distribution of the real SDSS observations. 

\begin{figure}[!t]
\resizebox{80mm}{!}{\includegraphics{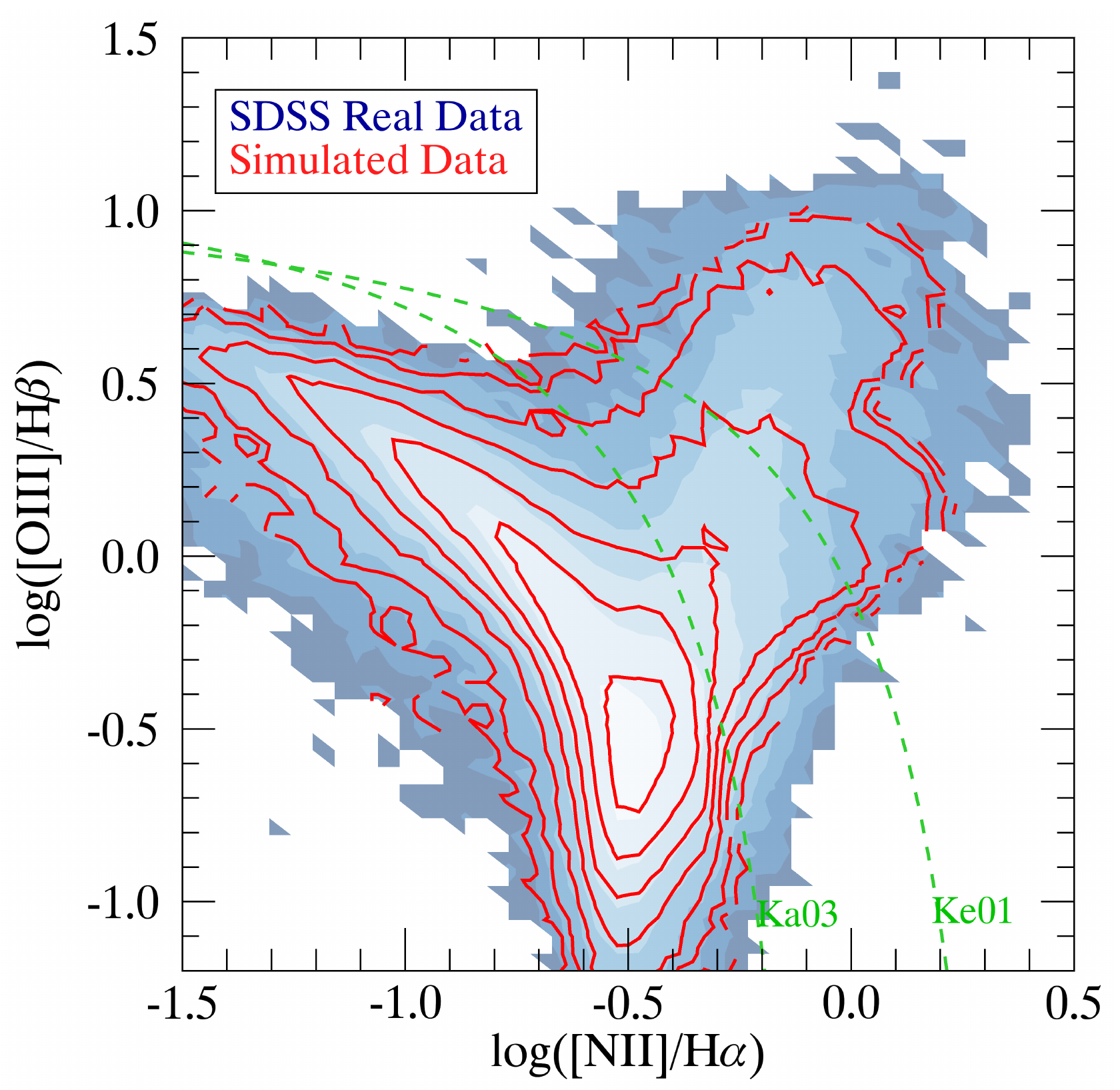}} \\
\caption{Our simulated sample of young galaxies in BPT space (red) plotted over the real sample of young galaxies from SDSS (blue). The distribution Schechter function was chosen so that the simulated sample would best fit the real data. \label{fig:bptover}}
\end{figure}

We compare our simulated set of galaxies to observations in three ways. We first examine our simulated sample and the observed galaxies in BPT space to compare the general shape of our sample distribution. As shown in Figure \ref{fig:bptover}, our sample is able to mimic the observed distribution of young galaxies with significant detection in all four BPT emission lines.

%%%%%%%%%%%%%%%%%%%%%%%%%%%%%%%%%%%%%%%%%%%%%%%%%
%%%%%%%%%%%%%%%%%%%%%%%%%%%%%%%%%%%%%%%%%%%%%%%%%

\begin{figure}[!t]
\resizebox{80mm}{!}{\includegraphics{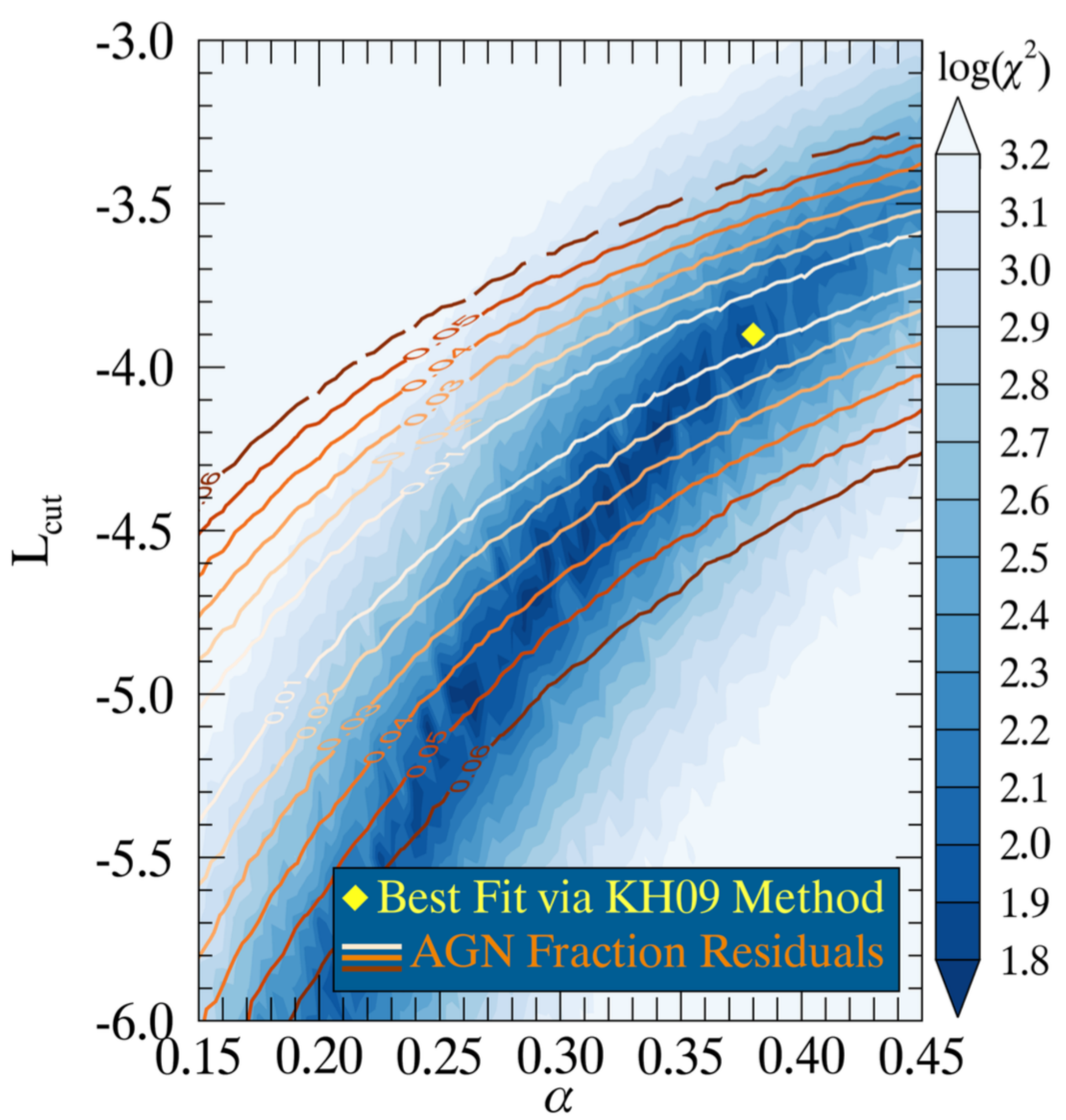}} \\
\caption{The observed-calculated Composite/AGN fraction residuals are shown in orange to white lines and overlaid over the blue $\chi^2$ values of the Eddington ratio distribution fit. Indicated as ``Best Fit'' from the \citetalias{KH09} method is the point of overlap with the minimum $\chi^2$ and residual fit with slope of $\alpha=$ 0.38, and lower cutoff of $L_{cut}=$ -3.9 (See Section \ref{sec:khsim}). \label{fig:chires}}
\end{figure}

We then check the output of our simulation by determining the fraction of galaxies that lie above the \citet{Kau03} (Ka03) and \citet{Kew01} (Ke01) demarcation lines. The observed occupation fractions above the demarcation lines are $14.7\%$ and $3.1\%$, respectively, for young galaxies with S/N$>3$ in all four BPT flux lines. We match the observed and simulated values by adjusting the simulated intrinsic Eddington ratio distribution, specifically the lower limit and $\alpha$. The occupation fraction residuals of these fits are shown in Figure \ref{fig:chires} as orange to white lines. As a further comparison, we calculate the Eddington ratio distribution of our simulated galaxies following the procedure discussed in Section \ref{sec:khsim}.

%%%%%%%%%%%%%%%%%%%%%%%%%%%%%%%%%%%%%%%%%%%%%%%%%
%%%%%%%%%%%%%%%%%%%%%%%%%%%%%%%%%%%%%%%%%%%%%%%%%

%%%%%%%%%%%%%%%%%%%%%%%%%%%%%%%%%%%%%%%%%%%%%%%%%
%SIMULATED EDDINGTON
%%%%%%%%%%%%%%%%%%%%%%%%%%%%%%%%%%%%%%%%%%%%%%%%%

\section{Determining the Eddington Ratio Distribution: Simulated SDSS Galaxies}\label{sec:khsim}

We apply the \citetalias{KH09} method to galaxies in our simulated sample that have emission line fluxes in all for BPT emission lines with a S/N$>$3, following the exact procedure used for the observed objects in Section \ref{sec:khreal}. The fraction of [\ion{O}{3}] due to the AGN component is similarly found for the entire set of simulated objects. The same two tracks from Section \ref{sec:khreal} have been selected to represent the AGN and LINER populations and the sample is split between these two ionization regimes using the same separation line (see Figure \ref{fig:rtrack}).

\begin{figure}[!h]
\resizebox{80mm}{!}{\includegraphics{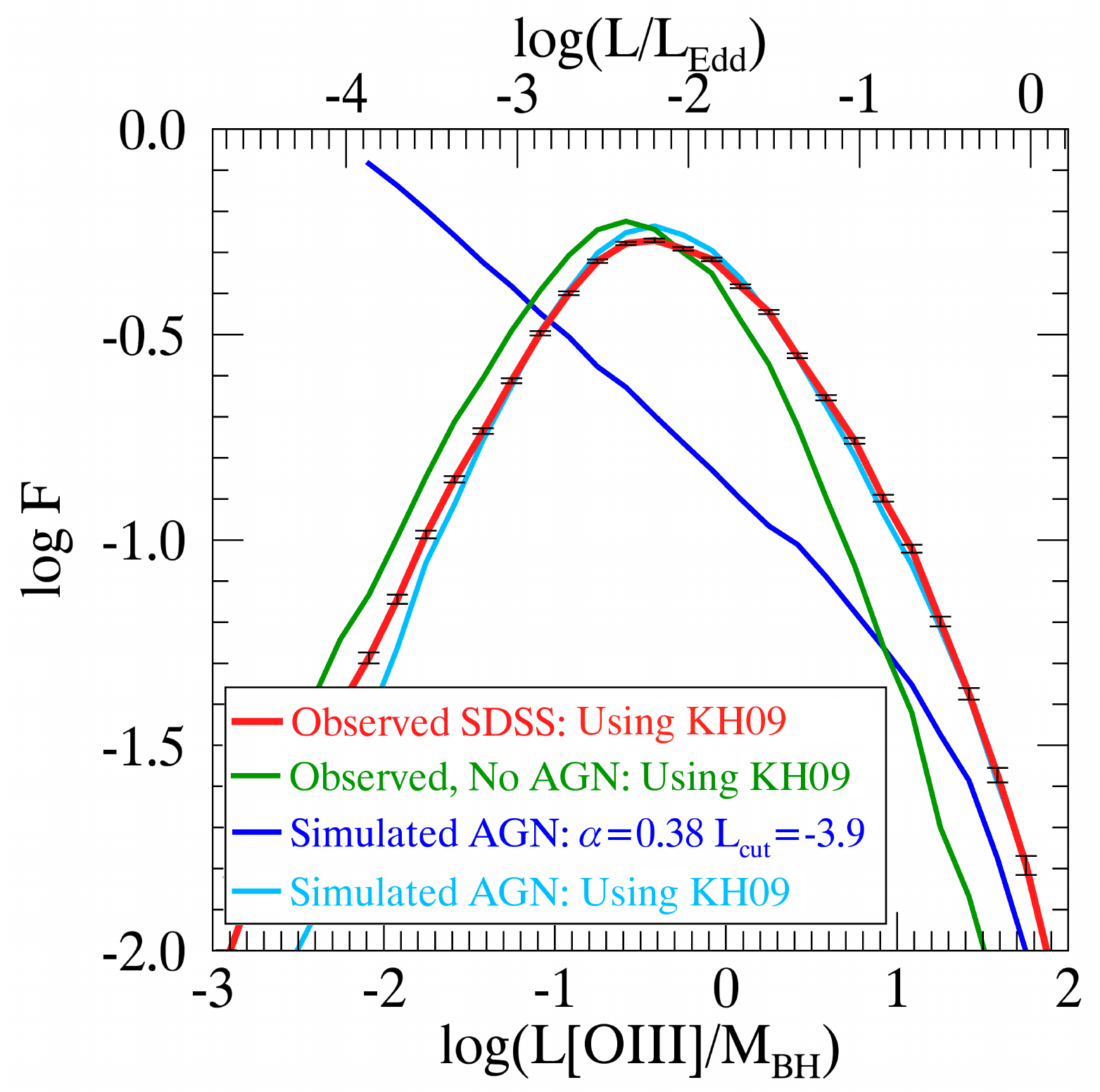}} \\
\caption{The Eddington ratio distribution extracted from the \citetalias{KH09} method for the observed SDSS data (red) appears lognormal, as well as the observed sample with only star-forming galaxies (No AGN; green). Our best fit fiducial model (dark blue) for the intrinsic Eddington ratio distribution is also shown. Note our intrinsic distribution is a power law with an exponential cutoff but becomes lognormal and consistent with the observed SDSS data after the \citetalias{KH09} method is applied (light blue).\label{rsedd}}
\end{figure}

Thus the ``observed'' contributions of AGN and star formation are computed for the simulated galaxies. It is useful to note that our simulated galaxies were built with a known AGN luminosity in [\ion{O}{3}] and are being compared to the AGN component determined empirically by this procedure. The differential Eddington parameter, as defined in Section \ref{sec:khreal}, is plotted in Figure \ref{rsedd}. \textit{This Eddington parameter distribution is approximately lognormal in shape and significantly different than the input intrinsic Eddington ratio distribution}.

%%%%%%%%%%%%%%%%%%%%%%%%%%%%%%%%%%%%%%%%%%%%%%%%%
%%%%%%%%%%%%%%%%%%%%%%%%%%%%%%%%%%%%%%%%%%%%%%%%%

After applying the \citetalias{KH09} method to our simulated data and the real SDSS data, we test the accuracy of our intrinsic function to reproduce the observed, real Eddington ratio distribution using $\chi^2$. The best fit is selected based on the overlap between the observed-calculated occupation fraction residuals from Section \ref{sec:mod} and the $\chi^2$ values of the fit to the real data, as shown in Figure \ref{fig:chires}. These parameters follow a similar trend, but a fit of $\alpha=$ 0.38 and lower cutoff $L_{cut}=$ -3.9 are the best overlap minimum for which the AGN fractions (defined above and below the Ka03 and Ke01 demarcations) are $14.7\%$ and $4.8\%$ respectively. This fit gives us 294,994 simulated young galaxies with S/N$>$3 in [\ion{O}{3}] which is comparable to the number of real young galaxies with the same signal strength, of which there are 294,934.

To test the uncertainty of our black hole mass calculations, we introduce an artificial scatter of 0.3 dex in log(M$_{BH}$) and run our simulation with our best fit Schechter function. When comparing our scattered Eddington ratio distribution to the best fit Eddington ratio distribution, we find that the recovered distributions agree within 0.03 dex in Eddington parameter space. With our best fit, we also find good agreement with the observed Eddington ratio distribution within 0.125 dex (see Figure \ref{rsedd}). When comparing the intrinsic to the observed Eddington ratio distribution, the most obvious difference is the turnover around -0.5 dex in $\log$(L[\ion{O}{3}]/$M_{BH}$). It is necessary to point out that when the \citetalias{KH09} method is applied to [\ion{O}{3}] detected star-forming galaxies, \textit{before the simulated AGN is added}, we extract an AGN contribution that is lognormal in shape with a similar turnover (Figure \ref{rsedd}). 

Thus the observed turnover is likely due to the intrinsic width of the star-forming sequence caused by observational and intrinsic scatter within the star-forming sequence and believed to be caused by the fluctuations and efficiency of accretion and star formation (e.g., \citealt{Kew01,Whi14}). The intrinsic width of the star-forming sequence contributes to hiding low luminosity AGN. A star-forming galaxy spectrum with emission line ratios that are offset from the selected ``pure'' star-forming loci will be found to have a significant AGN contribution, even if it has little or no AGN activity. This ``overestimates'' the AGN contribution for a substantial number of galaxies, such that this method cannot recover significant populations with small AGN contributions. Due to this scatter, any method to extract an AGN contribution from galaxy integrated fluxes in star-forming galaxies cannot yield a low AGN contribution.

%%%%%%%%%%%%%%%%%%%%%%%%%%%%%%%%%%%%%%%%%%%%%%%%%
% FRACTIONS
%%%%%%%%%%%%%%%%%%%%%%%%%%%%%%%%%%%%%%%%%%%%%%%%%
\section{A New Method for Determining the AGN Contribution by Position on the BPT Diagram}\label{sec:frac}

\begin{figure*}[!t]
\resizebox{180mm}{!}{\includegraphics{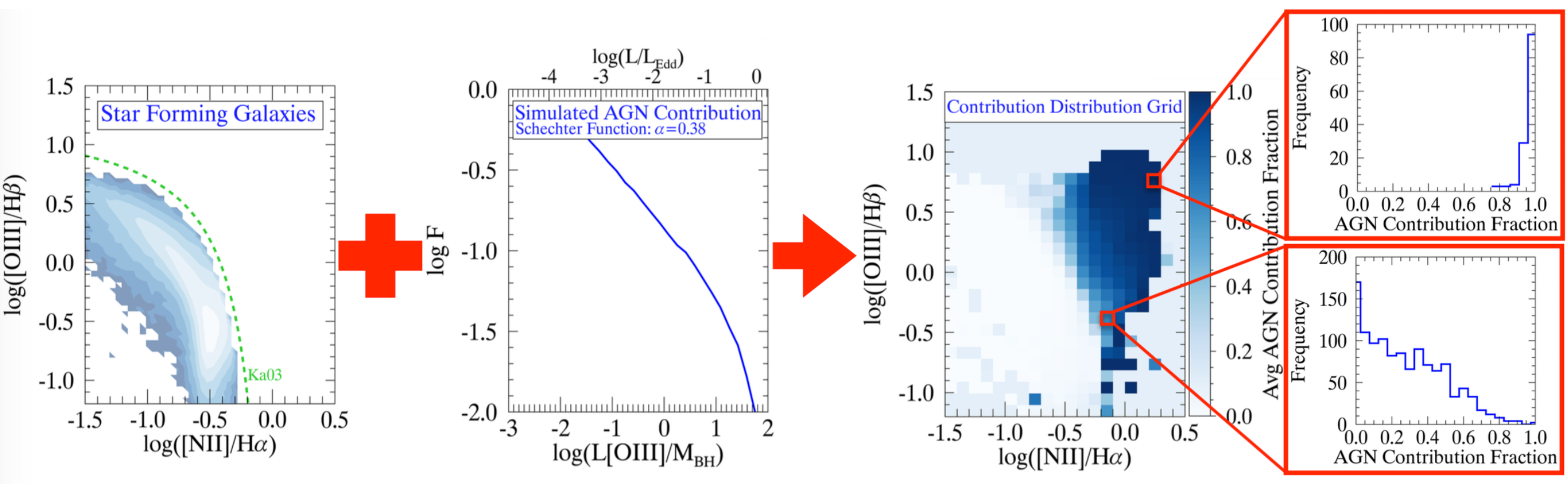}} \\
\caption{A schematic of our method for determining AGN contributions for an unknown data sample using a simulated Schechter function AGN distribution.\label{diagram}}
\end{figure*}

In this section, we develop a method for extracting the AGN contribution that accounts for the scatter of the star-forming sequence and preserves the probability distribution of finding AGN flux in each region on the BPT diagram with the aim of better reproducing the Eddington ratio distribution down to low values. Our procedure uses a simulated set of galaxies created following the process outlined in Section \ref{sec:khsim} a schematic of this method is given in Figure \ref{diagram}.

%%%%%%%%%%%%%%%%%%%%%%%%%%%%%%%%%%%%%%%%%%%%%%%%%
%%%%%%%%%%%%%%%%%%%%%%%%%%%%%%%%%%%%%%%%%%%%%%%%%

To recap, our simulation consists of adding an AGN luminosity component randomly drawn from a Schechter function distribution to a galaxy component selected as [\ion{O}{3}] undetected or star-forming based on the detection of [\ion{O}{3}] and/or position on the BPT diagram. We select the Schechter function parameters that produce a BPT diagram distribution consistent with the observations. Thus we have created a set of simulated galaxies with known AGN contributions across the BPT diagram. 

\begin{figure}[!h]
\resizebox{78mm}{!}{\includegraphics{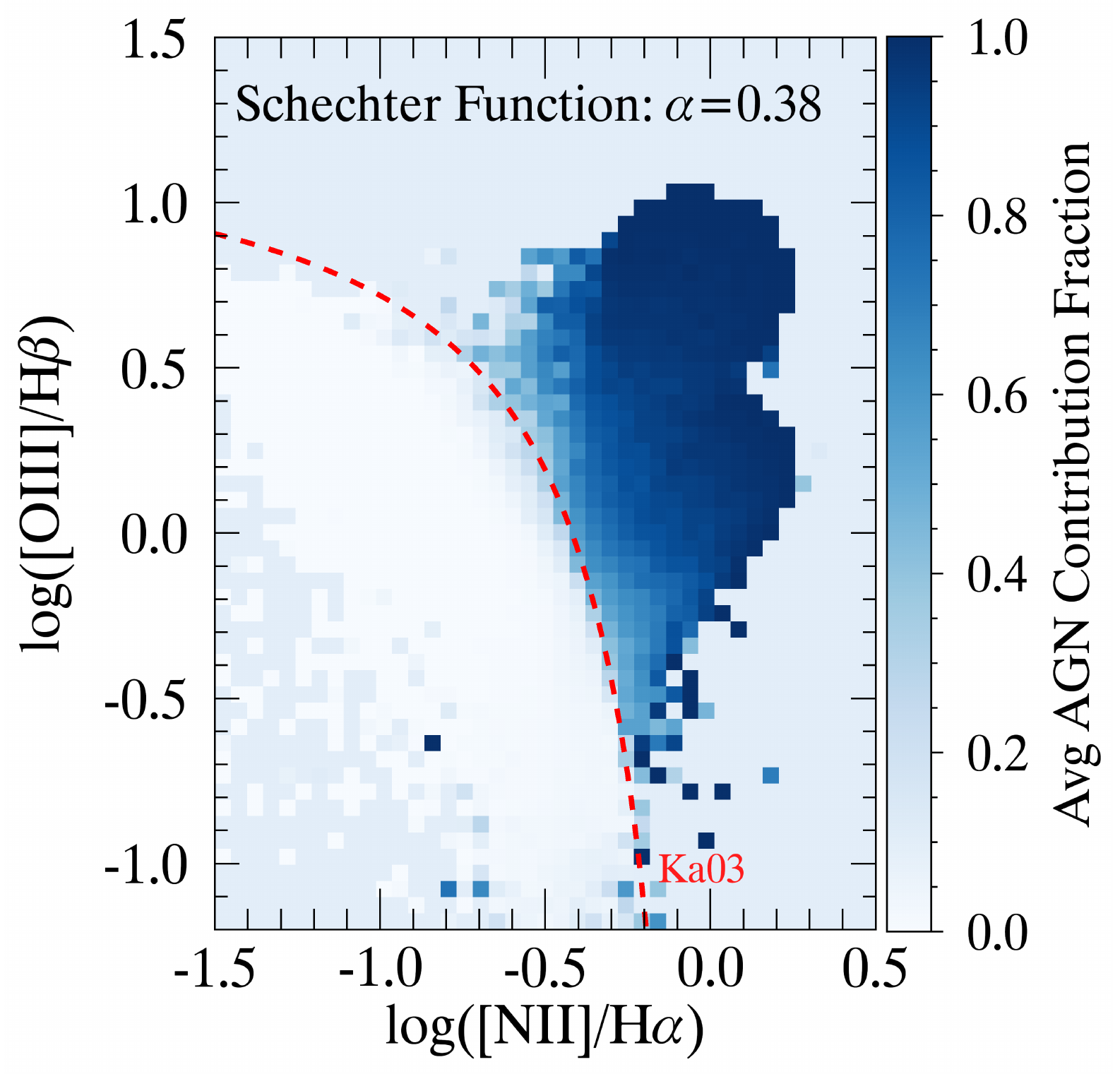}} \\
\resizebox{78mm}{!}{\includegraphics{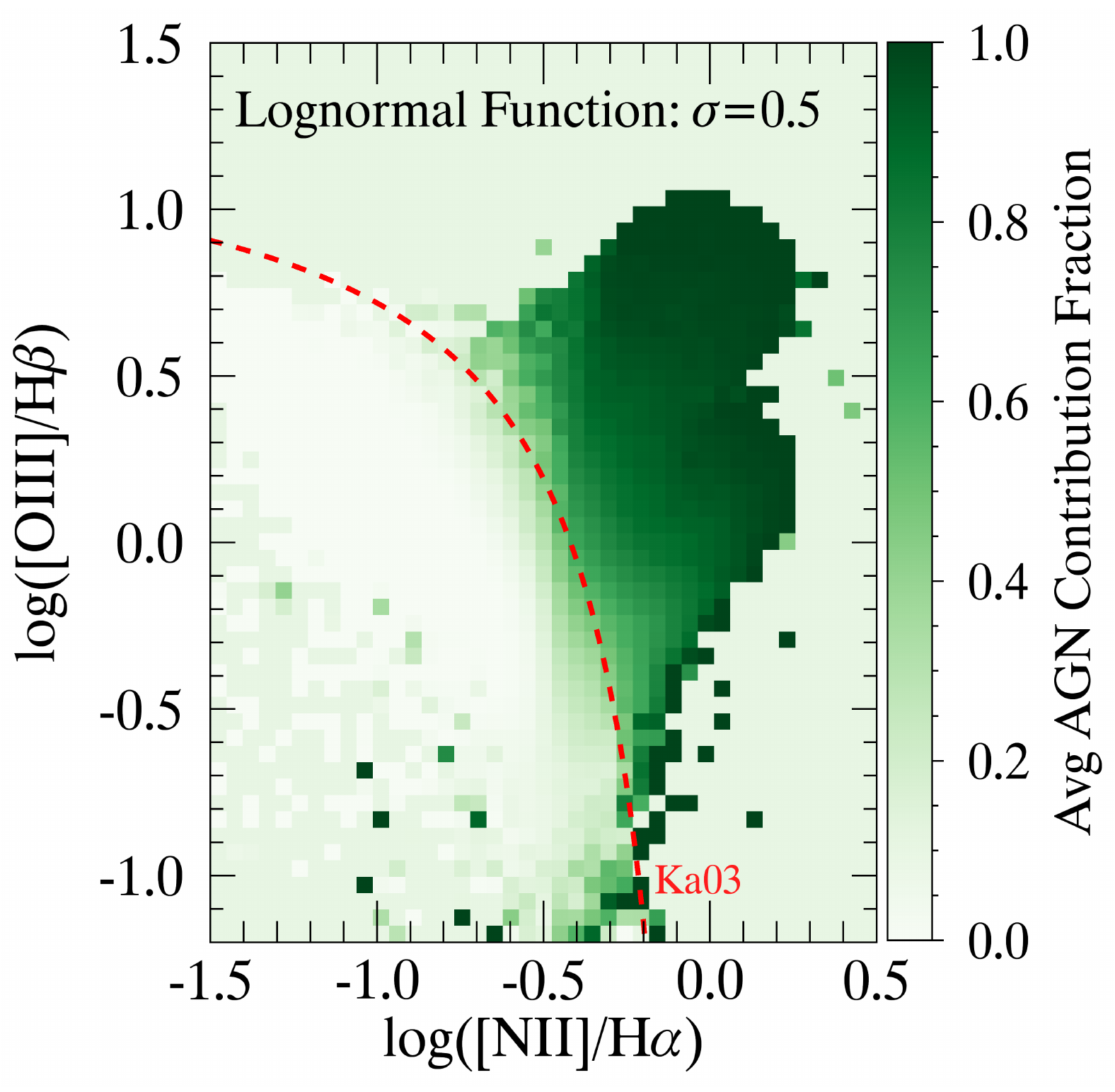}} \\
\caption{The BPT diagrams with the average AGN contribution fraction per grid square based on the distributions of our chosen grid functions, a simulated Schechter function (top) and lognormal function (bottom). \label{bptmet}}
\end{figure}

Using the positions of our simulated galaxy sample on the BPT diagram, we create a grid in BPT space in which each bin contains a sample of our simulated galaxies with the particular distribution of the AGN contribution fractions. This tells us the probability of a galaxy in each square having a specific AGN component. For a grid built from our Schechter function, the average AGN fraction in each of the selected bins is shown on the right side of the top panel of Figure \ref{bptmet}. While the average is depicted for simplicity, a distribution is computed for each bin so that the probability is conserved (Figure \ref{diagram}; right panels). As expected, the upper right portion of the BPT has the highest average AGN contributions to the total flux. We have thus built a tool from which we can extract the AGN contribution fraction based on the position of observed galaxies on the BPT diagram.

%%%%%%%%%%%%%%%%%%%%%%%%%%%%%%%%%%%%%%%%%%%%%%%%%
\begin{figure*}[!t]
\resizebox{180mm}{!}{\includegraphics{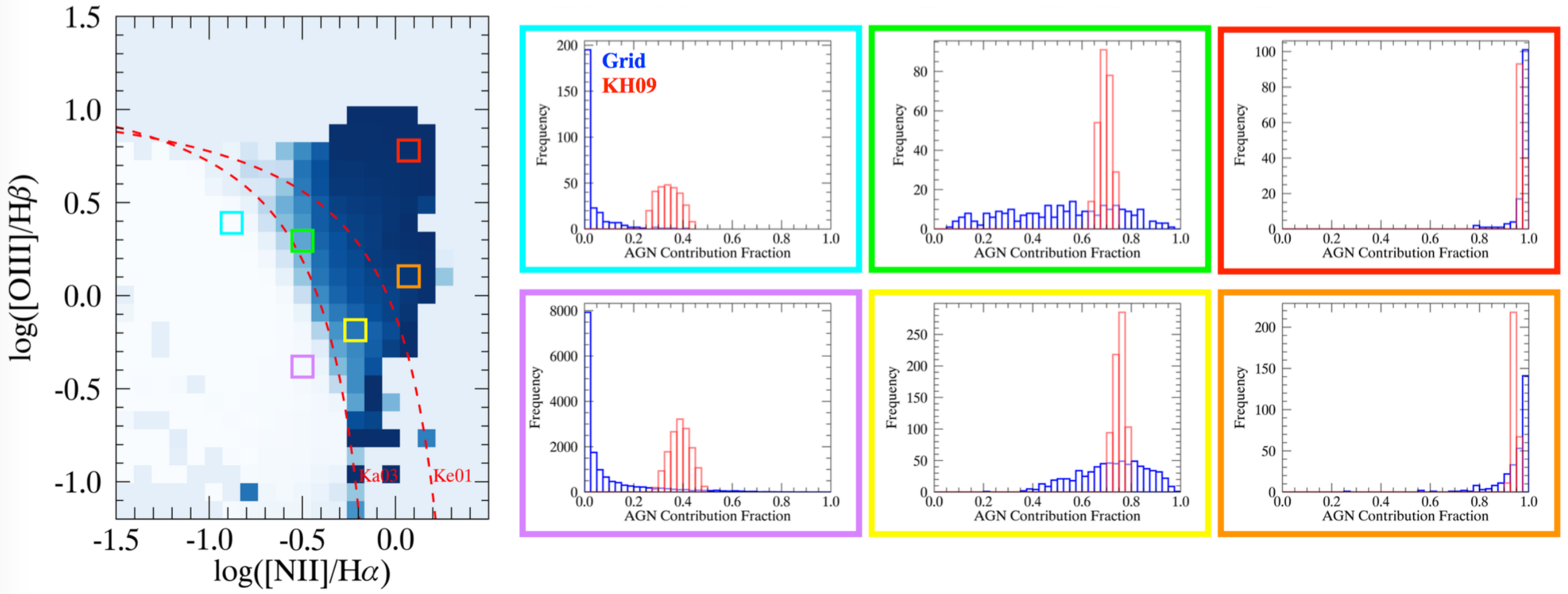}} \\
\caption{The grid method is able to more robustly extract an AGN contribution fraction based on a position on the BPT diagram. This method preserves the distribution in each grid region and has the ability to probe regions of small AGN contribution, accounting for the effects of dilution from star formation that can exist for low Eddington AGN in galaxies. These effects cannot be captured if one assigns a single AGN fraction for every position on the BPT diagram; rather, assigning a single AGN fraction based on a distance from a ``pure star-forming'' locus overestimates the AGN fraction for all galaxies but those in the AGN and LINER regions.  We show the AGN contribution fraction distributions for six select BPT regions for both the grid method (blue) and the KH09 method (red).
\label{dia:frac}}
\end{figure*}
%%%%%%%%%%%%%%%%%%%%%%%%%%%%%%%%%%%%%%%%%%%%%%%%%

We examine the AGN contribution fraction distributions across select regions of the BPT diagram corresponding to high and low ionization regions, as described in Section \ref{sec:mod} (Figure \ref{dia:frac}). This allows us to verify that the grid method is able to extract low AGN contributions even in regions where AGN emission may be diluted by star formation. In comparison, we show that by assigning a single AGN fraction for every position on the BPT diagram based on the distance from a ``pure star-forming'' locus, as in \citetalias{KH09}, we are unable to identify very small AGN contributions. We find that the grid method can more robustly extract a broad range of the AGN contribution to the flux while the \citetalias{KH09} method overestimates the AGN fraction for galaxies in the star-forming and composite regions in BPT space. This overestimation of the AGN fraction can account for the lognormal shape of the output Eddington ratio distribution and its large value at L/L$_{Edd}$ $\sim$ 0.01 that differ markedly from the input Schechter function. These differences can explain the discrepancy between the Eddington ratio distribution obtained by KH09 and those derived from X-ray surveys (see Figure \ref{fig:alledd}).

%%%%%%%%%%%%%%%%%%%%%%%%%%%%%%%%%%%%%%%%%%%%%%%%%
%%%%%%%%%%%%%%%%%%%%%%%%%%%%%%%%%%%%%%%%%%%%%%%%%

We divide the real SDSS data using this grid, separating our observed galaxies into bins based on the same grid for which our model gives us known AGN contribution fraction distributions. We assign each observed galaxy an AGN contribution fraction drawn randomly from the full distribution corresponding to that particular grid position on the BPT. Thus we extract AGN contributions to the flux of the real galaxies that preserve the distribution in AGN fraction within that region in BPT space. We follow the methods outlined in Section \ref{sec:khsim} to determine the ``observed'' Eddington ratio distribution.

To test this method and the completeness of the results, we chose a grid built from our ``best fit'' Schechter function and refer to it as our ``calibration'' function. We then use this grid to extract AGN contribution fractions from simulated data, assuming a known input models for the AGN Eddington ratio distribution: an intrinsic Schechter function distribution and an intrinsic lognormal distribution. Furthermore, we can fit an intrinsic distribution function to the extracted Eddington ratio distribution of the real data by applying our ``calibration'' grid to simulations built with our Schechter function with a large range in $\alpha$ and $L_{cut}$ as defined in Section \ref{sec:mod}. Using the calculated $\chi^2$ values, we develop a finer range of $\alpha$ and $L_{cut}$ to determine the best fit. This ``best grid fit'' is given by $\alpha=$ 0.40 and $L_{cut}=$ -3.75.

As an additional test, we can vary our calibration grid to be built from a lognormal function rather than our ``best grid fit'' Schechter function. The lognormal function we have elected to present is similar to what is found for young galaxies in \citetalias{KH09}. Additional lognormal functions were tested with varying slopes and centered at different Eddington ratios. These were only able to reproduce the observed low Eddington regime at the expense of under predicting the number of AGN at high Eddington ratios, and vice versa. We conclude that a single lognormal is not able to reproduce the full Eddington ratio distribution. Using two calibration grids allows us to verify how robust our technique is in recovering the input distribution and assess any potential biases due to the choice of ``calibration'' model. 

\begin{figure}[!t]
\resizebox{80mm}{!}{\includegraphics{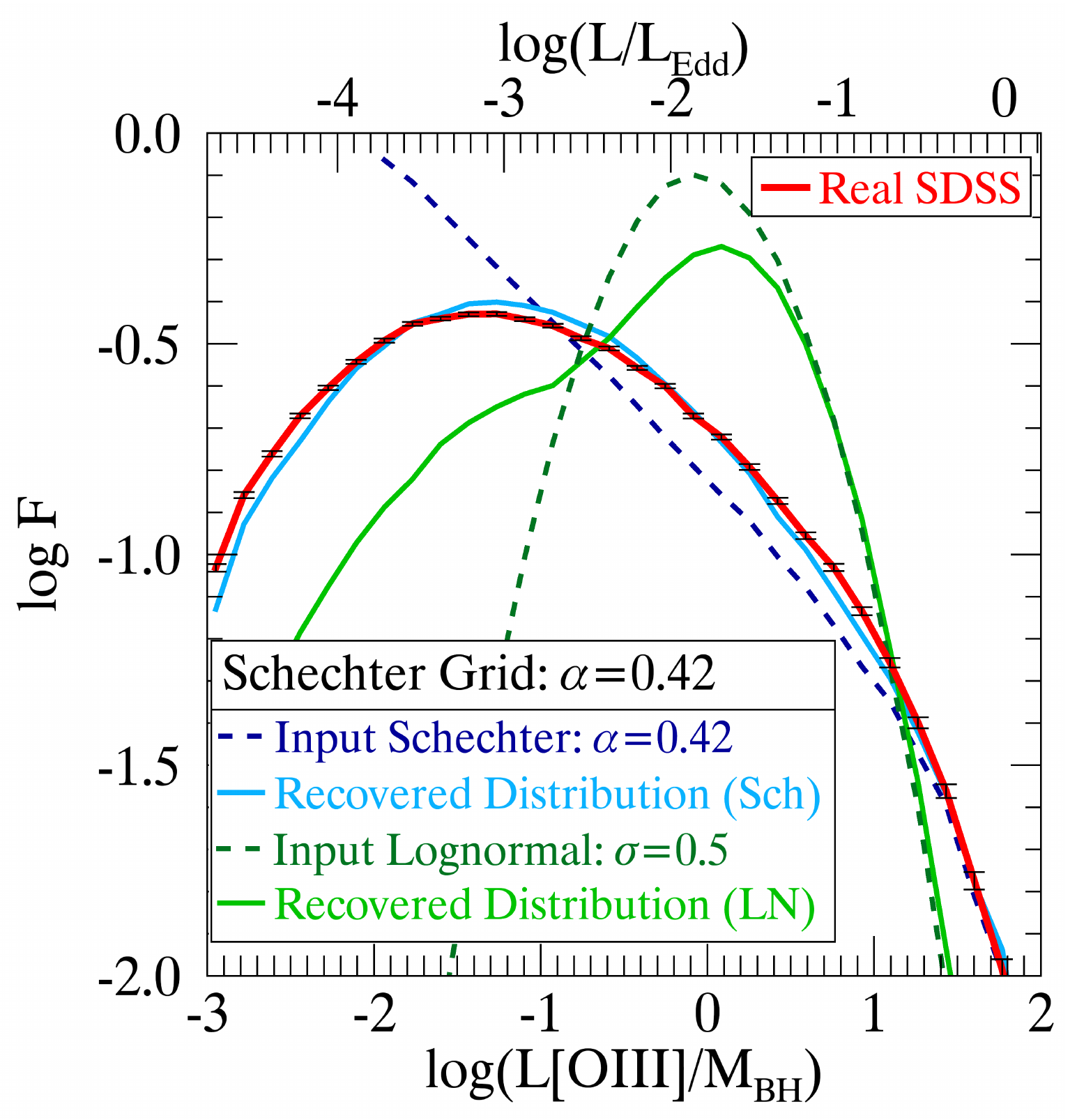}} \\
\resizebox{80mm}{!}{\includegraphics{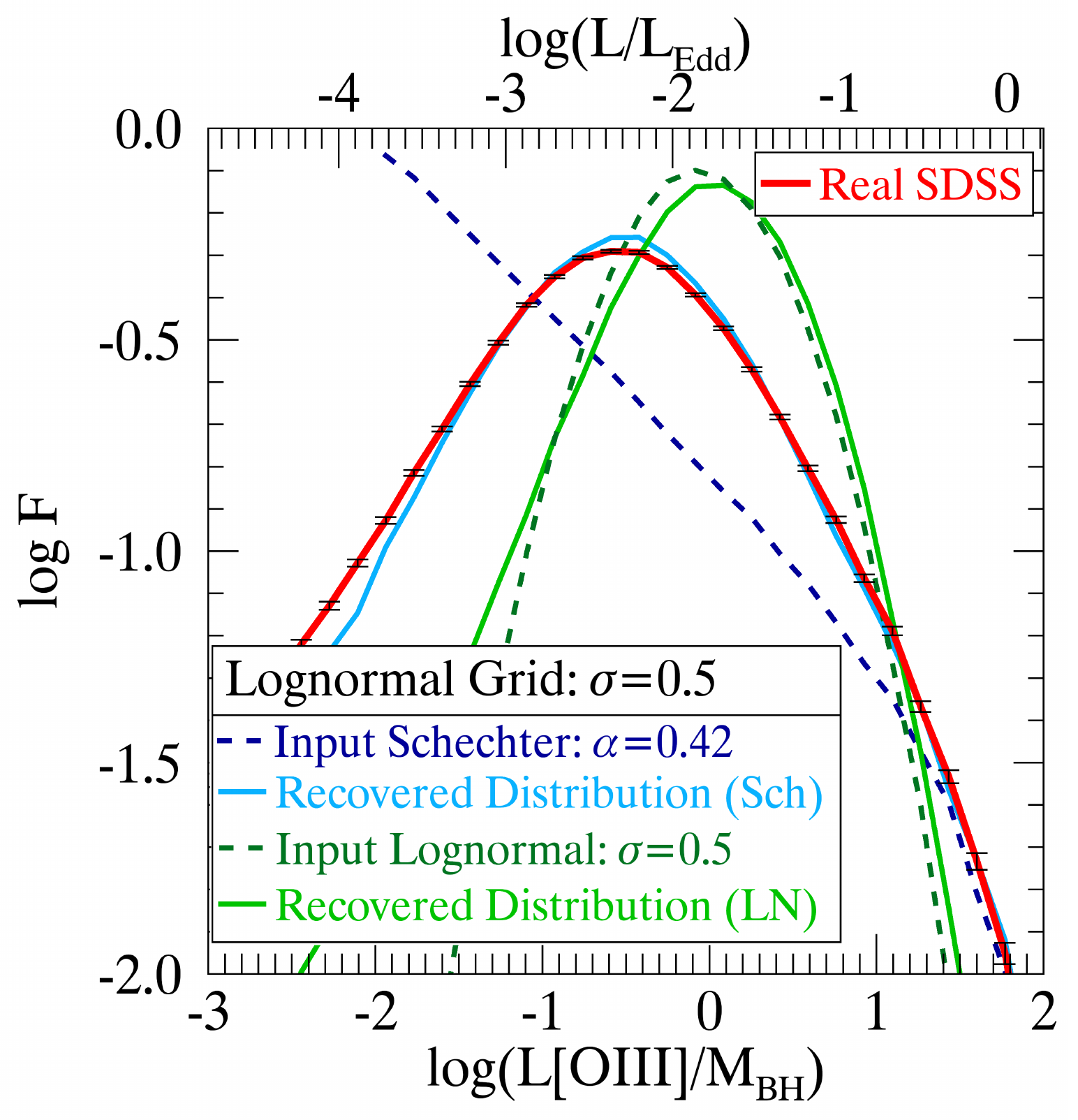}} \\
\caption{The Eddington ratio distribution for a Schechter function calibration grid with slope of $\alpha=$ 0.40 (dark blue, top) and a lognormal function calibration grid with slope of $\sigma=$ 0.5 (dark green, bottom). The extracted Eddington ratio for the real SDSS data is shown in red. The calibration grid was applied to two test distributions based on a Schechter function and lognormal function (dark blue, dark green respectively) to test the ability to recover those distributions. In both calibration grids, the real data more closely resembled the extracted intrinsic Schechter function (light blue) to -0.5 dex in $\log$(L[\ion{O}{3}]/$M_{BH}$). \label{calib}}
\end{figure}

The results of the following analysis are outlined below and referenced in Figure \ref{calib}. The top of this figure refers to the results of our first calibration grid selected, the ``best grid fit'' Schechter function with slope $\alpha$=0.40 and L$_{cut}$=-3.75 where the grid and input function are shown in dark blue.
\begin{itemize}	
\item We apply this grid to the real SDSS sample and extract an Eddington ratio distribution (red) that is similar to the calibration Schechter function (dark blue) in shape and exponential cutoff, if not amplitude, down to -0.5 dex in Eddington parameter.
\item This grid application is repeated with a sample of known intrinsic Eddington ratio distribution, one that is built from the same Schechter function used for the calibration (dark blue). We recover an Eddington ratio distribution (light blue) that matches the extracted distribution of the real SDSS sample (red) to within 0.07 dex in each bin of Eddington ratio.
\item As a further check, we change the intrinsic Eddington ratio distribution such that the simulated input is a lognormal function with width $\sigma=$ 0.5 (dark green). Our calibration grid extracts an Eddington ratio distribution (light green) which does not match the shape of the real Eddington ratio distribution until around -2 dex in Eddington ratio space, where the turnover to lower Eddington ratios exhibits a similar slope albeit different normalization.
\end{itemize}	

%%%%%%%%%%%%%%%%%%%%%%%%%%%%%%%%%%%%%%%%%%%%%%%%%
%%%%%%%%%%%%%%%%%%%%%%%%%%%%%%%%%%%%%%%%%%%%%%%%%

We then consider a different calibration function to test for potential bias in the grid creation process. We plot the results of our alternate calibration grid based on a lognormal Eddington ratio distribution with width $\sigma=$ 0.5 (the input function is shown in dark green) in the bottom of Figure \ref{calib}.
\begin{itemize}	
\item With this new grid, we extract the Eddington ratio distribution of the real SDSS sample (red) and find a wide peaked shape that does not match the input lognormal in slope, amplitude, or maxima.
\item We then apply the lognormal calibration grid to our ``best grid fit'' Schechter function (dark blue) and extract an Eddington ratio distribution (light blue) that again closely resembles the extracted Eddington ratio distribution of the real SDSS sample. They are consistent within 0.06 dex in each bin of Eddington ratio, to an Eddington ratio around -2 dex where they diverge slightly.
\item Once again, we switch the test function to a lognormal intrinsic distribution, using the same lognormal function as the calibration grid (dark green), and the grid is applied to this new distribution. The extracted lognormal Eddington ratio distribution (light green) is narrower and exhibits a slightly different maximum compared to the real data. 
\end{itemize}	
For both of the calibration grids, the extracted Eddington ratio distribution of the intrinsic Schechter function is consistent with the Eddington ratio distribution of the real data. These are then closely correlated in shape, if not amplitude, with the intrinsic Eddington ratio distribution to -0.5 dex in Eddington parameter. This technique is able to recover the shape of the Eddington ratio distribution to an order of magnitude lower in Eddington ratio than previously published methods.

\begin{figure}[!h]
\resizebox{80mm}{!}{\includegraphics{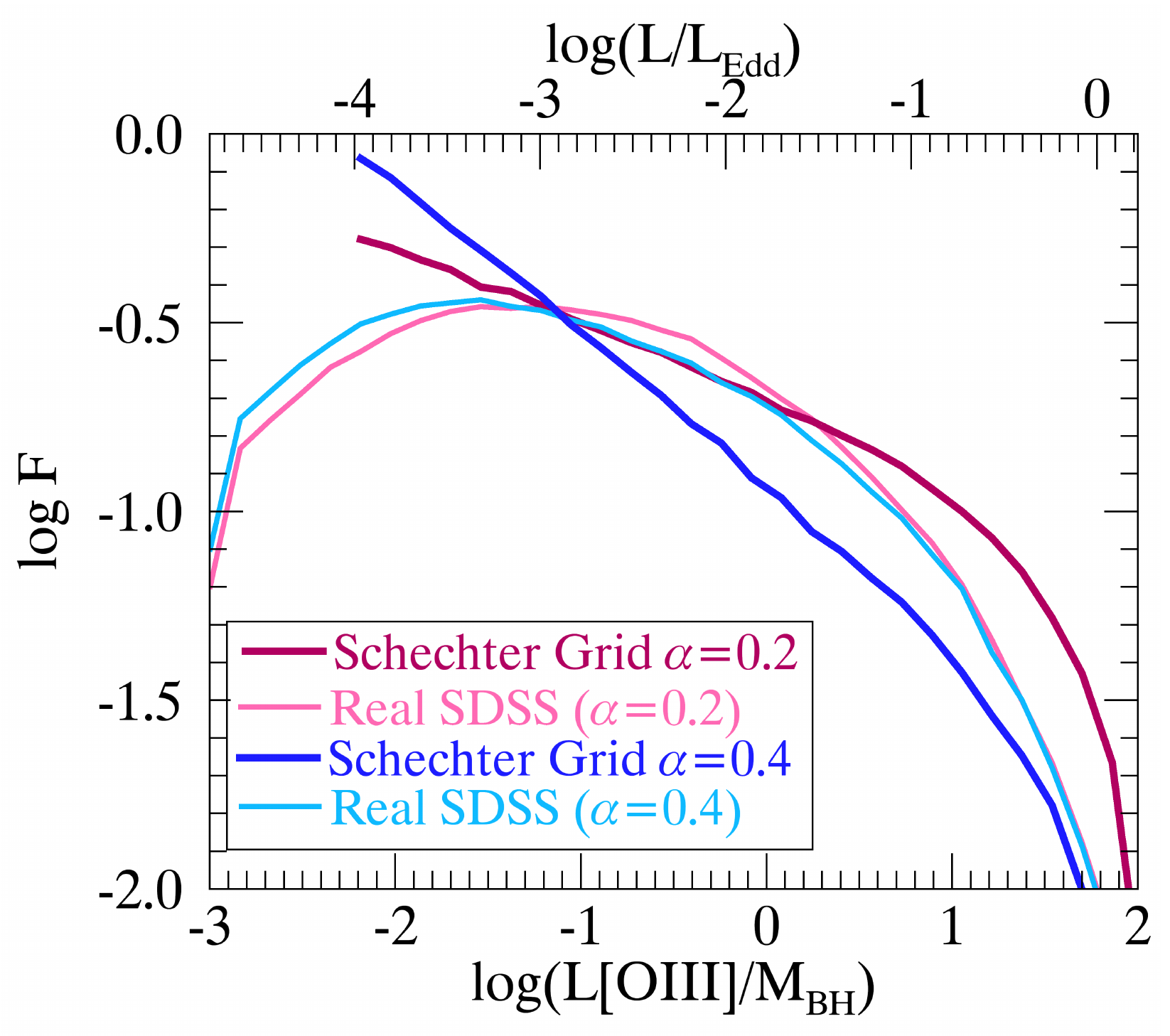}} \\
\resizebox{80mm}{!}{\includegraphics{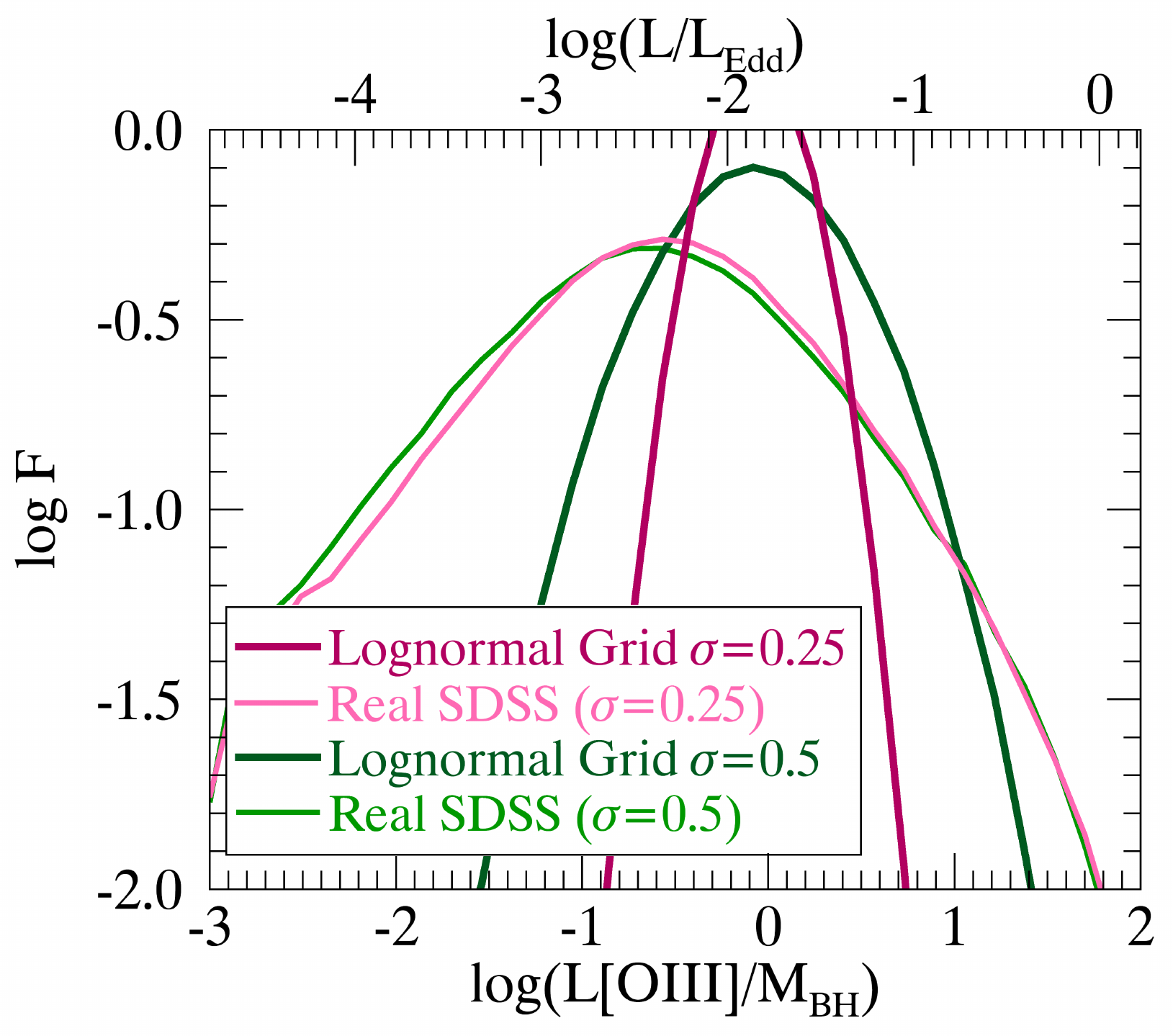}} \\
\caption{As a further test, we varied the slope of both calibration functions and found the extracted AGN contribution of the real data remained mostly unchanged within 0.25 dex, suggesting that once the calibration function is selected, varying its parameters does not introduce significant bias to the method.\label{sside}}
\end{figure}

We further explore the consequences of varying the slopes of the calibration functions on the extracted real data. The top of Figure \ref{sside} the Schechter function is shown with a slope of $\alpha=$ 0.2 and $\alpha=$ 0.4, while the bottom panel shows the lognormal with a width of $\sigma=$ 0.5 and $\sigma$= 0.25. As shown, we see that the extracted Eddington ratio distribution is consistent within 0.25 dex and 0.125 dex, respectively, despite changes to the input calibration distribution. For the Schechter function, it is important to note that the initial slope used in the fitting process with $\alpha=$ 0.38 was chosen purposefully based on the best fit to the distribution of the real data on the BPT diagram and more closely matches 0.4, whereas $\alpha=$ 0.2 is not as well matched. Thus, it is not unexpected that changing the calibration slope alters the peak of the extracted real Eddington ratio distribution as it deviates from the observed distribution.

%%%%%%%%%%%%%%%%%%%%%%%%%%%%%%%%%%%%%%%%%%%%%%%%%
%%%%%%%%%%%%%%%%%%%%%%%%%%%%%%%%%%%%%%%%%%%%%%%%%

We find that an intrinsic Schechter function is consistent with the Eddington ratio distribution of observed young SDSS galaxies. This method not only confirms that an intrinsic Schechter function is more likely to represent the Eddington ratio distribution than a lognormal distribution, but also accurately recovers the intrinsic shape of the Eddington ratio distribution to more than one order of magnitude in Eddington parameter than is possible using previously published methods.

%%%%%%%%%%%%%%%%%%%%%%%%%%%%%%%%%%%%%%%%%%%%%%%%%
% DISCUSSION
%%%%%%%%%%%%%%%%%%%%%%%%%%%%%%%%%%%%%%%%%%%%%%%%%
\section{Discussion and Summary}\label{sec:dis}

We use the galaxy spectroscopic catalog SDSS DR7 to investigate the intrinsic Eddington ratio distribution of galaxies obtained from optical spectroscopy.

We use the \citetalias{KH09} method of extracting AGN contributions from the BPT diagram on our sample of young observed galaxies to reproduce the lognormal Eddington parameter distribution found by \citetalias{KH09}. To test the consistency with our model, we create a simulated set of galaxies with an observed star-forming and simulated AGN component built from our fiducial model. To test our simulation, we compare it to the observational data by reproducing the general shape of the observed BPT diagram and population fractions. We apply the \citetalias{KH09} method to our young simulated sample, which returns a lognormal function consistent to the real observed young data. This suggests the recovered lognormal distribution could be an artifact of the method due to a bias against low luminosity AGN in star-forming galaxies.

Furthermore, we present a new method for determining the AGN contribution to the flux using the BPT optical diagnostic diagram. Our grid method confirms our earlier results that a Schechter function is a good description for the intrinsic Eddington ratio distribution of young galaxies. In addition, the method recovers the shape of our Schechter function to more than one order of magnitude in Eddington parameter than is possible using previously published methods.

Our best fit model to the AGN Eddington ratio distribution using the \citetalias{KH09} method is a Schechter function with exponential cutoff described by an $\alpha$ of 0.38 and a lower cutoff of -3.9. Figure \ref{fig:alledd} depicts a representation of this function compared to other popular models. The best fit model to the Eddington ratio distribution using the grid method is very similar with $\alpha$ of 0.40 and a lower cutoff of -3.75. Our best-fit slopes of the Schechter function ($\approx$0.4) are modestly flatter than those found by \citetalias{KH09} for passive galaxies ($\approx$0.6), similar to the findings of recent hydrodynamical simulations that find that galaxies with higher gas fractions have a flatter slope to the Eddington ratio distribution \citep{Gab13}.

For both methods, the extracted ``observed" Eddington ratio distribution from our simulated galaxies is comparable to the distribution obtained from the real data, suggesting that an intrinsic Schechter function distribution is still consistent with the ``observed'' lognormal distribution found by \citetalias{KH09}. If both the young and old galaxies have a Schechter function distribution, as suggested by X-ray observations (e.g., \citealt{Air13}), it is possible that AGN accretion is universal and the fueling mechanisms for all galaxies are much more comparable. Differences between individual AGN are most likely due to small scale stochastic variations rather than processes on galactic scales connected with star formation.

This work highlights the difficulties in identifying rapidly growing black holes in star-forming galaxies for any optical measurement with galaxy integrated fluxes due to the systematic bias against low luminosity AGN. Furthermore, it illustrates the extent to which emission from star formation in these host galaxies can dilute the observed AGN emission, whether it is a low luminosity AGN or an AGN in a particularly strong star-forming galaxy. It is especially difficult to separate the emission if the AGN flux is on scale with, or smaller than the flux from the star formation. This is in agreement with the results of \citet{Hop09,Gou09} and \citet{Tru15} in which host galaxy star formation and obscuration are found to impact the observed flux. Despite these difficulties, it successfully shows that a Schechter function Eddington ratio distribution is consistent with the SDSS observations.

In summary:
\begin{itemize}
\item An intrinsic Eddington ratio distribution for galaxies following a Schechter function is consistent with the observed lognormal distribution found by \citetalias{KH09} after accounting for the method of analysis.
\item Our grid method for extracting the AGN contribution further suggests that a Schechter function is a good description for the intrinsic Eddington ratio distribution of young galaxies.
\item We confirm the bias against observing and extracting a low to moderate luminosity AGN component in star-forming galaxies.
\item The Eddington distribution of young galaxies appears to be consistent with a universal power law with exponential cutoff, suggesting that the fueling mechanism between young and old galaxy populations may be more comparable and that differences in Eddington ratio are most likely due to small scale, stochastic variations rather than galaxy-scale processes.
\end{itemize}

%%%%%%%%%%%%%%%%%%%%%%%%%%%%%%%%%%%%%%%%%%%%%%%%%
\acknowledgments
We thank our collaborators as well as the anonymous referee for constructive comments that improved the paper. This work was supported in part by an NSF GK-12 Fellowship, Dartmouth Graduate Fellowship, as well as supported by the National Aeronautics and Space Administration under Grant Number NNX15AU32H issued through the NASA Education Minority University Research Education Project (MUREP) through the NASA Harriett G. Jenkins Graduate Fellowship activity. Funding for the SDSS and SDSS-II has been provided by the Alfred P. Sloan Foundation, the Participating Institutions, the National Science Foundation, the U.S. Department of Energy, the National Aeronautics and Space Administration, the Japanese Monbukagakusho, the Max Planck Society, and the Higher Education Funding Council for England. The SDSS Web Site is http://www.sdss.org/. We acknowledge the use of the Garching Value Added catalog and the Moustakas dust correction package. This research has made use of NASA's Astrophysics Data System.

%%%%%%%%%%%%%%%%%%%%%%%%%%%%%%%%%%%%%%%%%%%%%%%%%

\bibliography{ebs}

%%%%%%%%%%%%%%%%%%%%%%%%%%%%%%%%%%%%%%%%%%%%%%%%%

\end{document}